\documentclass[prd,aps]{revtex4}
\usepackage{graphicx}

\begin{document}

\title{Stability  properties of a formulation of Einstein's equations}

\author{Gioel Calabrese}
\email{gioel@lsu.edu}
\author{Jorge Pullin}
\email{pullin@lsu.edu}
\author{Olivier Sarbach}
\email{sarbach@phys.lsu.edu}
\author{Manuel Tiglio}
\email{tiglio@lsu.edu}

\affiliation{Department of Physics and Astronomy, Louisiana State
University, 202 Nicholson Hall, Baton Rouge, Louisiana 70803-4001}

\begin{abstract}

We study the stability properties of the Kidder-Scheel-Teukolsky (KST)
many-parameter formulation of Einstein's equations for weak
gravitational waves on flat space-time from a continuum and numerical 
point of view. At the continuum, performing a linearized analysis of
the equations around flat spacetime, it turns out that they have,
essentially, no non-principal terms. As a consequence, in the weak
field limit the stability properties of this formulation depend only
on the level of hyperbolicity of the system. At the discrete level
we present some simple one-dimensional simulations using the KST family.
The goal is to analyze the type of instabilities that appear as one
changes parameter values in the formulation. Lessons learnt in this
analysis can be applied in other formulations with similar properties.

\end{abstract}

\maketitle
\section{Introduction}

Numerical simulations of the Einstein equations for situations of
interest in the binary black hole problem do not run forever. The
codes either stop due to the development of floating point overflows,
or even if they do not crash, they produce answers that after a while
are clearly incorrect. It is usually very difficult to pinpoint a
clear reason why a code stops working.  Recently, Kidder, Scheel and
Teukolsky (KST) \cite{kst} introduced a twelve-parameter family of
evolution equations for general relativity. Performing an empirical
parameter study within a certain two-parameter subfamily, they were
able to evolve single black hole spacetimes for over 1000 M, where M is
 the mass of the black hole, something that had been very
difficult to achieve in the past.  It is of interest to try to understand
what makes some of the parameter choices better than others, in
particular given that a twelve dimensional parameter study appears
prohibitive at present. The intention of this paper is to take some
steps in this direction. We will first perform a linearized analysis
of the KST equations in the continuum, by considering small
perturbations around flat space-time. We will observe that the stability
of flat space-time is entirely characterized by the level of
hyperbolicity of the system. Since the latter is controlled by the
parameters of the family, this provides a first analytic guidance as
to which values to choose. Unfortunately, the result is somewhat weak, since it
just points to an obvious fact: formulations with higher level of
hyperbolicity work better.

In the second part of the paper we perform a set of simple numerical
tests.  We consider spacetimes where all variables depend on one
 spatial coordinate, which we consider compactified for simplicity, and
time. We are able to exhibit explicitly the various types of
instabilities that arise in the system.  Some of the results are
surprising. For the situation where the system is weakly hyperbolic,
the code is strictly non-convergent, but it might appear to converge
for a significant range of resolutions. We will see that the addition
of dissipation does not fix these problems, but actually can
exacerbate them. It is often the case in numerical relativity that
discretization schemes that are convergent for strongly hyperbolic
equations are applied to weakly hyperbolic formulations. The examples
of this section will teach us how dangerous such a practice is and
confirm the analytic results of reference \cite{convergence}. This
part of the paper is also instructive in that the KST system has only
been evolved with pseudo-spectral methods. We use ordinary integration
via the method of lines.

The plan of this paper is as follows. In the next section we will
discuss several notions of stability that are present in the
literature, mostly to clarify the notation. In section III we discuss
the stability of the KST equations in the continuum under linearized
perturbations. In section IV we discuss the numerical simulations.

\section{Different definitions of stability}

The term {\it stability} is used in numerical relativity in several
different ways. We therefore wanted to make the notation clear at least
in what refers to this paper.  Sometimes the notion of stability is
used in a purely analytic context, while some other times it is used
in a purely numerical one.  Within analytical contexts, there are
cases in which it is used to mean well posedness, as in the book of
Kreiss and Lorenz \cite{kreiss1}. In such a context well posedness means that the
norm of the solution at a fixed time can be bounded by the norm of the
initial data, with a bound that is valid for all initial data. In
other cases it is intended to measure the growth of perturbations of a
certain solution within a formulation of Einstein's equations, without
special interest in whether the equations are well posed or not.

At the numerical level, a scheme is sometimes defined as stable if it
satisfies a discrete version of well posedness. This is the sense in
which stability (plus consistency) is equivalent to convergence via
the Lax theorem \cite{richtmeyer}. Examples of this kind of
instability are present in the Euler scheme, schemes with Courant factor that are
too large, or other situations where the amplification factor (or its
generalization) is bigger than one.  Finally, the term stability has also
historically been loosely used to mean that a simulation runs for a
certain amount of time before crashing, or that the errors remain
reasonably bounded as a function of time for a certain amount of time.

At the continuum, part of the problem we want to look at can then be stated as
follows. Suppose there is a certain solution $u_0$ of Einstein's equations which is
bounded for all times. One can
prescribe the initial data $f_0$ that uniquely determines, through the use of the
evolution equations, that solution. However, if
one gives initial data slightly different from $f_0$, the
corresponding solution might grow without bound or even blow up in a finite
amount of time. This can happen either because this is
the physical situation, or because the new solution $u_1$ does not satisfy the
constraints, or does satisfy them but it is in a gauge that is
becoming ill defined. 
Numerical simulations may blow up  even
when one tries to model physically stable spacetimes. This is the case
for evolutions of 
Schwarzschild or Kerr black holes
(which are known to be at least linearly stable with respect to
physical perturbations.)

The time growth of solutions of a given set of partial differential
equations can be 
classified according to whether or not it can be bounded by a time function that
does not depend on the initial data. This leads
to the concept of well posedness. Namely, a system of
partial differential equations is said to be well posed 
at a solution $u_0(t)$ if there is a $T$ and some norm such
that
\begin{equation}
||u(t)|| \leq F(T) ||u(0)||  \; \; \mbox{for} \; 0\leq t\leq T
\label{bound}
\end{equation}
for all solutions $u(t)$ with $u(0)$ sufficiently close to $u_0(0)$.
For linear equations the bound can be tightened to an
exponential, but for nonlinear systems $F(T)$ might be any  other
function that can even diverge at finite $T$. The strength of
equation (\ref{bound}) characterizes the property that the bound is
independent of the
details of the 
initial data; in particular, of their frequency components. This is
important in numerical 
simulations, where more (and higher) frequency components appear
when resolution is increased. Well posedness is a necessary condition
in order to have long term, convergent evolutions, but it is not 
sufficient. Indeed, the function  $F(T)$ 
can grow quickly in time. In order to control $F(T)$ one has to go beyond
well posedness and study non-principal terms. We will now apply these
ideas to the KST equations.

\section{The KST family of equations and their linear stability}
\label{kst}

\subsection{The formulation}

Starting from the 
Arnowitt, Deser and Misner (ADM) equations, KST 
derive a family of strongly hyperbolic first-order evolution equations
for the three-metric ($g_{ij}$), the extrinsic curvature ($K_{ij}$), and the
spatial derivatives of the three-metric ($d_{kij} \equiv \partial_k
g_{ij}$), which generalizes previous well posed formulations \cite{fr,ec}.
A priori prescribing  the densitized
lapse $\exp{(Q)}$ as a function of spacetime, the lapse $N$ is given by
\begin{equation}
N = g^\sigma e^Q, \label{densitizing_lapse}
\end{equation}
where $g$ is the
determinant of the three-metric and $\sigma$ a parameter. 
The shift vector, $N^i$, is also assumed to be prescribed as a
function of spacetime. We define the derivative operator
along the normal to the hypersurfaces as
\begin{displaymath}
\partial_0 = \frac{1}{N} (\partial_t - \pounds_{\vec{N}}).
\end{displaymath}
The vacuum evolution equations have the form
\begin{eqnarray}
\partial_0 g_{ij} &=& - 2K_{ij}\, , \label{Eq:KST1}\\
\partial_0 K_{ij} &=& - \frac{1}{2} \partial^k d_{kij}
  + \frac{1}{2}\partial^k d_{(ij)k} + \frac{1}{2} g^{ab} \partial_{(i} d_{|ab|j)}
  - \left(\frac{1}{2} + \sigma \right) g^{ab}\partial_{(i} d_{j)ab}
\nonumber\\
 &+& \zeta\, g^{ab} C_{a(ij)b}
  + \gamma\, g_{ij} C - e^{-Q} \partial_i\partial_j (e^Q) + {\cal R}_{ij}\, ,
\label{Eq:KST2}\\
\partial_0 d_{kij} &=& - 2\partial_k K_{ij}
  + \eta\, g_{k(i} C_{j)} + \chi\, g_{ij} C_k + {\cal R}_{kij}\, ,
\label{Eq:KST3}
\end{eqnarray}
where the constraint variables
\begin{eqnarray}
& C = \frac{1}{2} g^{ab}\partial^k (d_{abk} - d_{kab}) + {\cal R}, & \hbox{(Hamiltonian constraint)}
\nonumber\\
& C_j = \partial^a K_{aj} - g^{ab}\partial_j K_{ab} + {\cal R}_j, & \hbox{(momentum constraint)}
\nonumber\\
& C_{kij} = d_{kij} - \partial_k g_{ij}\, , & \hbox{(definition of $d_{kij}$)}
\nonumber\\
& C_{lkij} = \partial_{[l} d_{k]ij}\, . & \hbox{(''closedness'' of $d_{kij}$)}
\nonumber
\end{eqnarray}
have been added to the right-hand side of the evolution equations
with some free parameteres $\zeta, \gamma , \eta, \chi$.
Here, $\partial^k \equiv g^{kl} \partial_l\,$ and
$b_j$, $d_k$ are defined as the traces of $d_{kij}\,$:
\begin{displaymath}
b_j = g^{ki} d_{kij}\, , \;\;\;
d_k = g^{ij} d_{kij}\, .
\end{displaymath}
The ``remainder'' terms ${\cal R}$, ${\cal R}_i$, ${\cal R}_{ij}$ and ${\cal R}_{kij}$
are homogeneous polynomials of degree $2$ in $d_{kij}$, $\partial_i Q$
and $K_{ij}$ (this will have direct consequences on the linear
stability, as discussed below), where the
coefficients may depend on $g_{ij}$.
Finally, the Lie derivative of the symbols $d_{kij}$ is
\begin{displaymath}
\pounds_{\vec{N}} d_{kij} = N^l\partial_l d_{kij} + d_{lij}\partial_k N^l
 + 2d_{kl(i}\partial_{j)} N^l + 2 g_{l(i} \partial_{j)}\partial_k N^l.
\end{displaymath}

The evolution system has characteristic speeds
$\{ 0, \pm 1, \pm\sqrt{\lambda_1}, \pm\sqrt{\lambda_2}, \pm\sqrt{\lambda_3} \}$, 
where
\begin{eqnarray}
\lambda_1 &=& 2\sigma,
\nonumber\\
\lambda_2 &=& 1 + \chi - \frac{1}{2}(1 + \zeta)\eta + \gamma(2 - \eta + 2\chi),
\nonumber\\
\lambda_3 &=& \frac{1}{2}\,\chi + \frac{3}{8}(1 - \zeta)\eta
     -\frac{1}{4}(1 + 2\sigma)(\eta + 3\chi). \nonumber
\end{eqnarray}
It should be noted that these are the characteristic speeds with
respect to the $\partial _0$ operator. This means that $\lambda =0,1$
correspond to propagation along the normal to the hypersurfaces or the
light cone, respectively. The characteristic speeds $\mu$ with respect to
the $\partial_t$ operator are obtained from these after the
transformation
\begin{displaymath}
\mu \mapsto N\mu + N^i n_i\; ,
\end{displaymath}
where $n^i$ is the direction of the corresponding characterisitic mode.

The conditions under which the system is completely ill posed (CIP),
weakly hyperbolic (WH) or strongly
hyperbolic (SH) (see next subsection for these definitions) were found 
in KST. The system is CIP if any of the above speeds is complex,
while it is WH if $\lambda_j \geq 0$ for $j=1,2,3$ but one of the
conditions (\ref{eq:SH1}, \ref{eq:SH2}) below is violated.
Finally, the system is SH provided that \cite{st}
\begin{eqnarray}
&& \lambda_j > 0, \;\;\; \hbox{for $j=1,2,3$}, \label{eq:SH1}\\
&& \lambda_3 = \frac{1}{4}(3\lambda_1 + 1) \;\;\;
  \hbox{if $\lambda_1 = \lambda_2$}. \label{eq:SH2}
\end{eqnarray}
For example, if the parameters ($\zeta$, $\gamma$, $\eta$,
$\chi$) are all zero the dynamics is equivalent to the ADM equations 
 written in first order form with fixed densitized lapse and fixed
shift (which are WH). If $\sigma =0$ as well, then the system is
equivalent to the ADM equations with fixed lapse and shift (which are
also WH). 

In KST seven extra parameters are introduced and used to make
changes of variables in $K_{ij}$ and $d_{kij}$. When performing this
change of variables the constraint $C_{kij}=0$ is also used in order to
trade spatial derivatives of the three-metric for $d_{kij}$.
Thus, the equations with the new variables have different
solutions off the constraint surface.
However, one can see that at the linear level
this change of variables does not involve the addition of
constraints.

\subsection{Linear stability}

Here we study the linear stability of the KST system considering perturbations
of  Minkowski spacetime written in Cartesian coordinates.  That is, we assume 
that the background quantities are
\begin{displaymath}
g_{ij} = \delta_{ij}\, , \;\;\;
K_{ij} = 0, \;\;\;
d_{kij} = 0, \;\;\;
N^i = 0, \;\;\;
Q = 0.
\end{displaymath}
In fact, since both $Q$ and $N^i$ can be freely specified, we will first fix
them to their background values for simplicity (the more general case
is analyzed at the end of this section.)  Because the non-principal part 
of the evolution equations for $K_{ij}$ and $d_{kij}$ depends quadratically
on $K_{ij}$, $d_{kij}$ and the derivatives of $Q$, only the principal part
remains when we linearize these evolution equations.
More precisely, the linearized equations 
have the following structure:
\begin{eqnarray}
\dot{g}_{ij} &=& -2 K_{ij}\, , \nonumber\\
\dot{K}_{ij} &=& (\mbox{\boldmath $A$} d)_{ij}\, , \label{Eq-LinKST}\\
\dot{d}_{kij} &=& (\mbox{\boldmath $B$} K)_{kij}\, , \nonumber
\end{eqnarray}
where $\mbox{\boldmath $A$}$ and $\mbox{\boldmath $B$}$ are spatial, first order, differential linear 
operators with constant coefficients. Furthermore, perturbations of
the three-metric do not appear in the
equations for $K_{ij}$ and $d_{kij}$. 
As a consequence, it is sufficient to consider only these equations;
after having solved  them, the three-metric is obtained through a time
integration. So,
the relevant linear evolution equations have the form
\begin{equation}
\dot{u} = \sum_{j=1}^3 A^j \partial _j u \; , 
\label{four}
\end{equation}
where $u$ is a ``vector'' formed by the components of $K_{ij}$ and $d_{kij}$
($u$ has, thus, $24$ independent components); and $A^j$ are 
$24 \times 24$ matrices.

Since the matrices $A^j$ have constant coefficients and we want to make
an analysis in the absence of boundaries, we consider a three-torus with periodic 
boundary conditions as a domain and analyze equation (\ref{four})
through  Fourier expansion. That is, we write
\begin{displaymath}
u(t,\vec{x}) = \sum_{\vec{k}} \hat{u}(t,\vec{k}) e^{i \vec{k}\cdot
\vec{x}} \;.
\end{displaymath}
The solution to equation (\ref{four}) is, then
\begin{displaymath}
u(t,\vec{x}) = \sum_{\vec{k}} e^{i P(\vec{k}) t} e^{i \vec{k}\cdot{\vec{x}}}\,
\hat{u}(0,\vec{k}) \; , 
\end{displaymath}
where $P(\vec{k})$, called the symbol, is defined by
\begin{displaymath}
P(\vec{k}):=  \sum_{j=1}^{3}{A^j k_j } \; .
\end{displaymath}
There are three different possibilities for the behavior of $e^{i P(\vec{k}) t}\,$:
\begin{enumerate}
\item The system is SH: $P$  has real
eigenvalues and is diagonalizable, the solution simply oscillates in time.
Nevertheless, we should keep in mind that the three-metric can grow
linearly in time if $\vec{k} = \vec{0}$, see Eqs. (\ref{Eq-LinKST}).
It is not difficult to see that this mode corresponds to an infinitesimal
coordinate transformation.

\item  The system is only WH: $P$ has real eigenvalues but is not
diagonalizable. Besides the zero-frequency growth in the metric, the
Jordan blocks of dimension $n+1$ in $P$ allow for  growth in $u$ that
goes as $(k t)^n$, where $k=|\vec{k}|$.

\item The system is CIP: $P$ has at least one complex
eigenvalue. Besides the previous growing modes, there can be
exponential ---frequency dependent--- growth, i.e.~
$u \approx \exp{(\mbox{constant}\times k  t)}$.
\end{enumerate}

To summarize, the linear stability properties of Minkowski spacetime depend 
only on the hyperbolicity of the system, since the non-principal terms
are zero (except for the evolution equation for $g_{ij}$,
but this equation decouples to linear order.)
Stability advantages of the choice of parameters empirically found by
KST in their analysis of a single black hole spacetime cannot be
displayed in the example we have considered. 

Finally, we study the effect of having a non-vanishing 
 linearized (maybe densitized) lapse and shift:
In this case, there are two extra terms in the evolution equation
for $K_{ij}$ and $d_{kij}\,$:
\begin{eqnarray}
\dot{K}_{ij} &=& (\mbox{\boldmath $A$} d)_{ij} - \partial_i\partial_j Q , \nonumber\\
\dot{d}_{kij} &=& (\mbox{\boldmath $B$} K)_{kij} + 2\partial_k\partial_{(i} N_{j)} .
\label{Eq-LinKSTLB}
\end{eqnarray}
Since we assume that $Q$ and $N^i$ are prescribed, the
solutions to these equations will have the form of the sum of
a homogeneous solution and a particular solution. It is not
difficult to find a particular solution, because we expect that a
variation of $Q$ and $N^i$ induces an (infinitesimal) coordinate
transformation $x^\mu \mapsto x^\mu + X^\mu$ with respect to
which, in this case, 
\begin{eqnarray}
g_{ij} &\mapsto& g_{ij} + 2\partial_{(i} X_{j)}, \nonumber\\
K_{ij} &\mapsto& K_{ij} - \partial_i\partial_j X^t\, ,\label{Eq:GaugeTrsf}\\
d_{kij} &\mapsto& d_{kij} + 2\partial_k\partial_{(i} X_{j)}. \nonumber
\end{eqnarray}
In fact, if we make the ansatz $K_{ij} = -\partial_i\partial_j f$,
$d_{kij} = 2\partial_k\partial_{(i}\xi_{j)}$, we get a particular solution to
the system (\ref{Eq-LinKSTLB}) if
\begin{eqnarray}
\dot{f} &=& 2\sigma \partial^l\xi_l + Q, \nonumber\\
\dot{\xi}_j &=& \partial_j f + N_j\, . \nonumber
\end{eqnarray}
After a Fourier decomposition, this system has the form
\begin{displaymath}
\dot{v}(t,\vec{k}) = i \mbox{\boldmath $M$}(\vec{k}) v(t,\vec{k}) + q(t,\vec{k}),
\end{displaymath}
where $v$ and $q$ are formed by the Fourier amplitudes of
$f$, $\xi_j$ and $Q$, $N _j$, respectively $(j=1\ldots3)$,
and
\begin{displaymath}
\mbox{\boldmath $M$}(\vec{k}) =
\left( \begin{array}{cccc}
  0 & 2\sigma k_1  & 2\sigma k_2 & 2\sigma k_3 \\
k_1 & 0 & 0 & 0 \\
k_2 & 0 & 0 & 0 \\
k_3 & 0 & 0 & 0 
\end{array} \right) \; .
\end{displaymath}
By choosing the homogeneous solution to satisfy the appropriate initial
data, it is enough to consider  a particular solution with $v(t=0,\vec{k})=0$,
\begin{displaymath}
v(t,\vec{k}) =    e^{i \mbox{\boldmath $M$}(\vec{k}) t} \int\limits_0^{t} e^{-i
\mbox{\boldmath $M$}(\vec{k}) s}
 q(s,\vec{k}) ds  .
\end{displaymath}
If $\sigma > 0$ (as must be the case if the system is SH), the
matrix $\mbox{\boldmath $M$}$ is diagonalizable, with real eigenvalues $0$, 
$\pm\sqrt{2\sigma |\vec{k}|^2}$.
Since the norm of $\exp({i \mbox{\boldmath $M$}(\vec{k})t})$ can be bounded by a
constant that does not depend on $t$ nor on $\vec{k}$ (for example,
if $2\sigma = 1$, the matrix $\mbox{\boldmath $M$}(\vec{k})$ is symmetric
and $\exp{(i \mbox{\boldmath $M$}(\vec{k})t})$ is unitary), this implies that
$v$ can grow at most linearly in time if $q$ is uniformly bounded in time.

If $v$ grows at most linearly in time,
the same will hold for the main variables $g_{ij}$, $K_{ij}$ and
$d_{kij}$, see (\ref{Eq:GaugeTrsf}). Again, we do not see any
dependence on the parameters found by KST in this particular example.

\section{Numerical experiments}
\label{num_section}


In \cite{convergence} we show analytically that
the iterated Crank--Nicholson (ICN) method with any number of
iterations
 and the second-order
Runge-Kutta methods do not yield convergent discretization schemes for
systems of equations that are not strongly hyperbolic, even if
dissipation is added. Here we
exemplify this through the numerical evolution of  the KST system in 
a simple situation.  We evolve
(\ref{Eq:KST1},\ref{Eq:KST2},\ref{Eq:KST3})
in one dimension. That is, all quantities in those equations are
assumed to depend only on $t$ and $x$.

 The sense in which stability is used in this section is that of the 
Lax equivalence theorem, i.e. a scheme is said to be
stable if the numerical solution 
$u^n_k$ at time $t=n \Delta t$ satisfies
$$
||u^n|| \leq f(t) ||u^0||
$$
for all initial data $u^0_k$ and small enough $\Delta t$ and $\Delta
x$. Here the index $n$ corresponds to the time step and $k$ to the
spatial mesh point.  It is in this sense that stability plus
consistency is equivalent, through Lax's theorem,  to convergence 
(convergence with respect to
the same discrete norm in which the scheme is stable and
consistent). Note that in this context consistency means that the
discrete equation  approaches the continuum one in the limit of 
zero grid spacing. In what follows we
use the discrete norm
$$
||u^n||^2 = \sum_k \left(u^n_k \right)^2 \Delta x \; .
$$

We will use second order Runge-Kutta (RK) for the time evolution. The
spatial derivatives were discretized with centered differences, adding
explicit lower order numerical dissipation as well, the amount of it
being arbitrary. Most of the simulations presented here will be for
RK, though the results are very similar for ICN (see the end of the
section).  In order to isolate our simulations from effects coming
from boundary conditions, we will choose, as in Section \ref{kst},
periodic boundary conditions.  Also, even though the analytic
stability analysis of Section \ref{kst} and of Ref. \cite{convergence}
is linear, here we will present {\em nonlinear} simulations, but of
situations that are close to flat spacetime. In all cases we will use
the same initial data, same numerical scheme, and change only the
continuum formulation of Einstein's equations.

Writing the evolution equations as
$$
\dot{u} = Au' + l.o. \; , 
$$
where ``l.o.'' stands for lower order terms and time integration 
  done with RK corresponds to 
\begin{equation} 
u^{n+1}_k = \left[ 1 + C \left( 1 + \frac{C}{2}\right) \right] u^n_k \,,
\label{RK}
\end{equation}
where 
$$
C = \frac{A \lambda }{2} \delta_0 - I\tilde{\sigma} 
\lambda \delta^4 + \Delta t \times l.o. \; , 
$$
with $\lambda = \Delta t/\Delta x$ the Courant factor,
$\tilde{\sigma}$ the dissipation parameter and $I$ the identity
matrix, and 
\begin{eqnarray}
\delta_0 u_k &=& u_{k+1}-u_{k-1} \; ,\\
\delta^4 u_k &=& u_{k+2}-4u_{k+1} + 6u_k -4u_{k-1} +u_{k-2} \;.
\end{eqnarray}

Our strategy for comparing numerical stability with the level of 
hyperbolicity is the
following: we fix all parameters of the KST system except one, and change that single
parameter in such a way that one gets a SH, WH or CIP system.

The metric that we evolve, giving the corresponding initial
data, is
\begin{equation}
ds^2 = e^{f_1+f_2}(-dt^2 + dx^2) + dy^2 + dz^2 \;,   \label{plane_metric1}
\end{equation}
where $f_1=f_1(t+x)$ and $f_2=f_2(t-x)$ are arbitrary functions. This 
is an exact solution of Einstein's equations that is pure gauge, i.e.~it
is flat. In order to illustrate our point it is enough to make a
simple choice for these functions, such as
$f_1 = A\sin{(x+t)}$ and $f_2 = 0$ with $x\in[-\pi,\pi]$; 
furthermore, for the simulations shown below we choose $A=0.01$.

\subsection{Densitizing the lapse}

A necessary condition in order to get well posedness within the KST
system is to choose a positive densitization for the lapse, $\sigma
>0$. Furthermore, one has to take $\sigma
=1/2$ if one wants physical characteristic speeds. It is worth
noticing that with $\sigma =1/2$ and the metric 
(\ref{plane_metric1}), the perturbation in the
densitization of the lapse $Q$ is zero; thus, as shown in Section
\ref{kst}, even off the constraint surface the solutions to SH
formulations should only oscillate in time, except for possible
zero-frequency linear growth. 

In a similar way, choosing $\sigma =0$ results in a WH 
system, and $\sigma < 0$ in a CIP one. 

Densitizing the lapse is not enough for well posedness. One also has to add
the constraints to the evolution equations. For definiteness,
in this subsection we will show simulations using
$\zeta=-1, \gamma=0, \eta=4, \chi=0$
(which  corresponds to the EC formulation without 
making the change of variables), but the main conclusions do not depend on this
particular choice.

\subsubsection{The SH case ($\sigma = 1/2$)} 

\begin{figure}[ht] 
\begin{center} 
\includegraphics*[height=6cm]{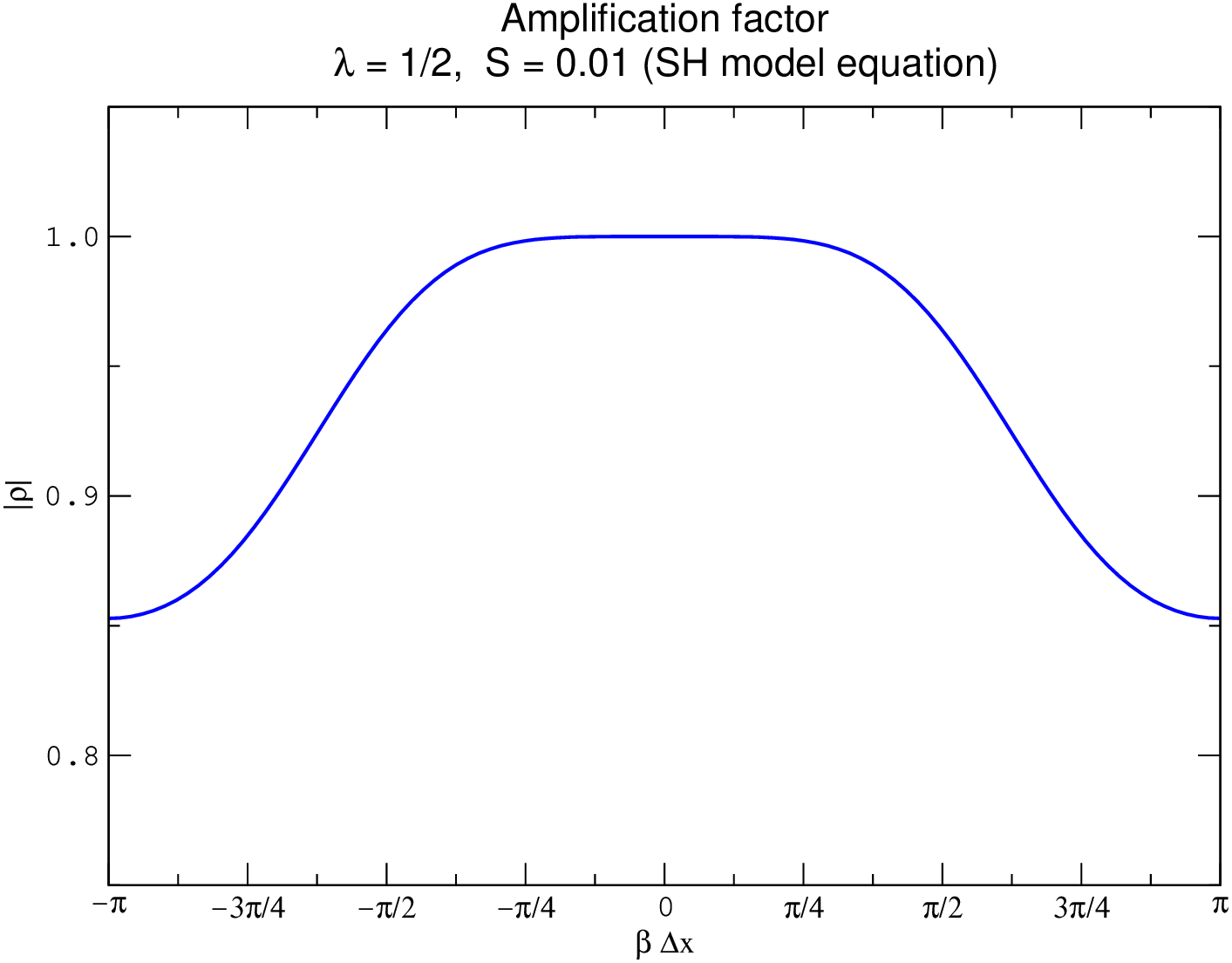}
\includegraphics*[height=6cm]{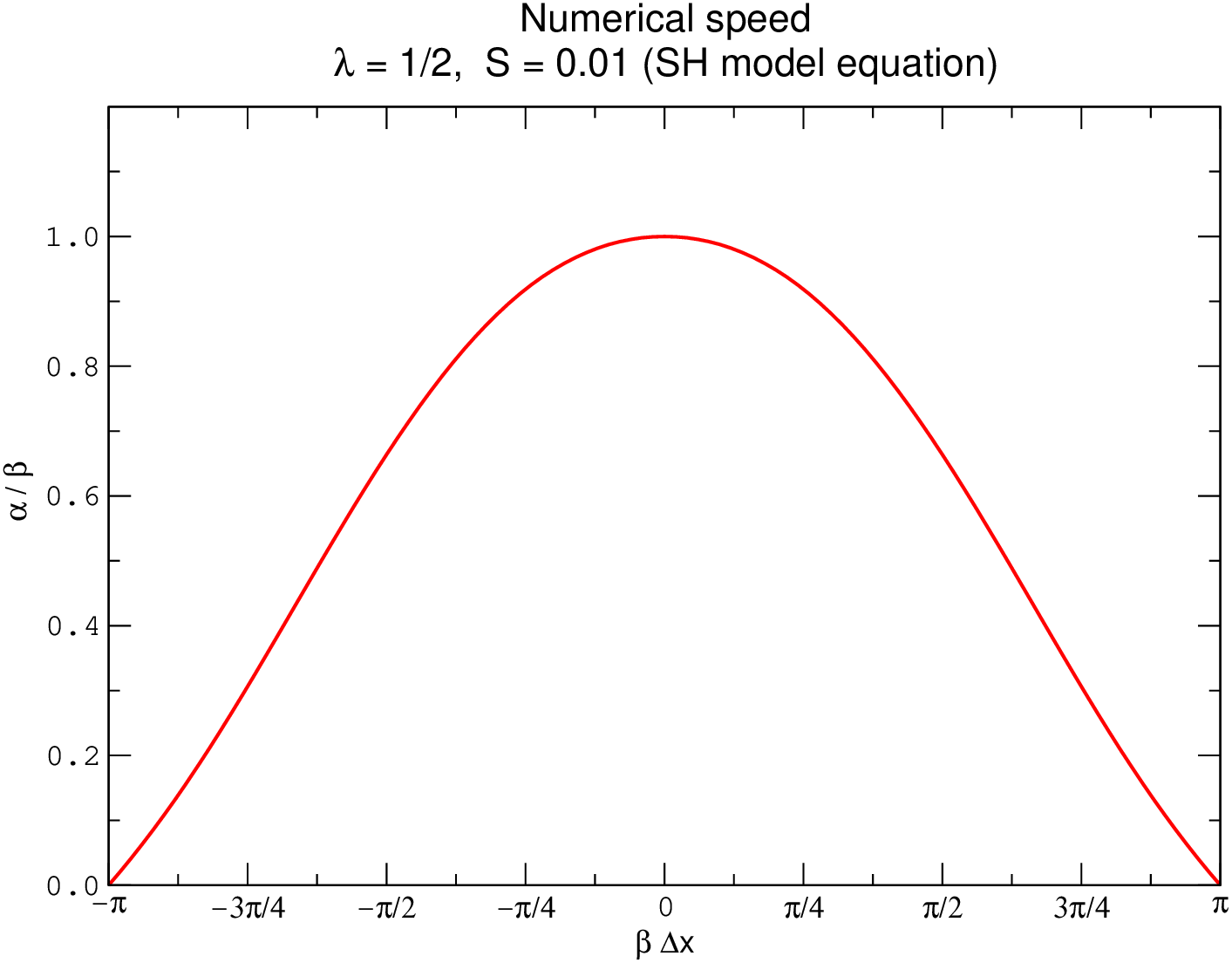}
\caption{Plot of the amplification factor and the numerical speed
associated with difference scheme (\ref{RK}).   If the product $\beta
\Delta x$ is small enough, then the damping and the error in the
propagation speed of the correspondent Fourier mode are very small. \label{dispdissip}}
\end{center}
\end{figure}

In order to understand some of the features of the SH numerical
 results that we will present we start analyzing, from a discrete point of view, the same
 numerical scheme when applied to the model equation 
\begin{equation}
\dot{u} = u' \;. \label{model_eq}
\end{equation}
Inserting the discrete Fourier mode
\begin{equation}
u^n_k = \hat{u}(\omega,\beta) e^{i(\omega n \Delta t + \beta k \Delta x)} \label{Fmode}
\end{equation}
into the difference scheme (\ref{RK}), 
one obtains the discrete dispersion relation
\begin{equation}
e^{i\omega \Delta t} = 1 + \left(i\lambda \sin(\beta \Delta x)-
16S\sin^4\frac{\beta \Delta x}{2}\right)\left(1 +
 i\frac{\lambda }{2} \sin(\beta \Delta x) -8S\sin^4\frac{\beta \Delta
 x}{2}\right)
 = \rho(\beta \Delta x) \,, \label{eiomegat}
\end{equation}
where  $S = \tilde{\sigma} \lambda$. Equation (\ref{eiomegat}) should
be seen as a relation between $\omega $ and $\beta $.   
The discrete symbol, $\rho(\beta \Delta x)$ tells us how the different Fourier
 modes are damped and dispersed (see, e.g. \cite{thomas}, chapter 7).  Writing 
$\omega = \alpha + i b$, we have that
\begin{eqnarray}
e^{-b\Delta t} &=& | \rho(\beta \Delta x) | \,, \qquad
\mbox{(amplification factor)} \label{ampfactor}\\
\frac{\alpha}{\beta} &=& \frac{1}{\beta \Delta t} \arg (\rho(\beta
\Delta x)) \,.
 \qquad \mbox{(numerical speed)}\label{disc_speed}
\end{eqnarray}
Choosing, for example, $\lambda =1/2$ and $S=0.01$, one can see
\cite{convergence} that the scheme (\ref{RK})
for Eq.(\ref{model_eq}) is stable and, therefore, convergent. From now on
 we will use, unless otherwise stated, these values. Figure
 \ref{dispdissip} shows the 
magnitude of the amplification factor (\ref{ampfactor}) and the
numerical speed (\ref{disc_speed}) for this
 choice of parameters.

\begin{figure}[ht] 
\begin{center} 
\includegraphics*[height=6cm]{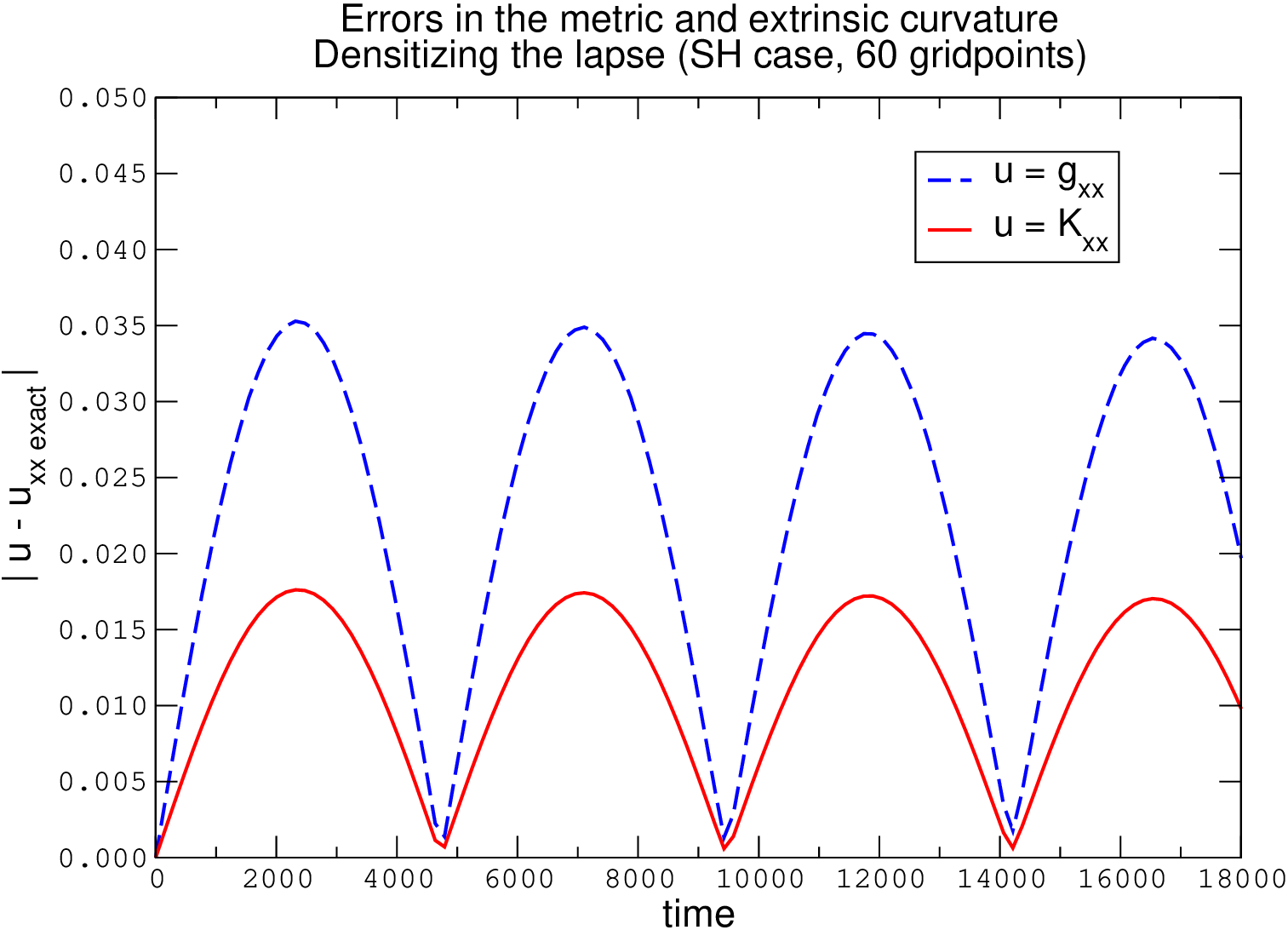}
\caption{$L_2$ norm of the errors for the metric and extrinsic
curvature in the SH case. These errors are mostly from numerical speed
difference, as described in the text. \label{errors_sh_dens_gK}}
\end{center}
\end{figure}

\begin{figure}[ht] 
\begin{center} 
\includegraphics*[height=6cm]{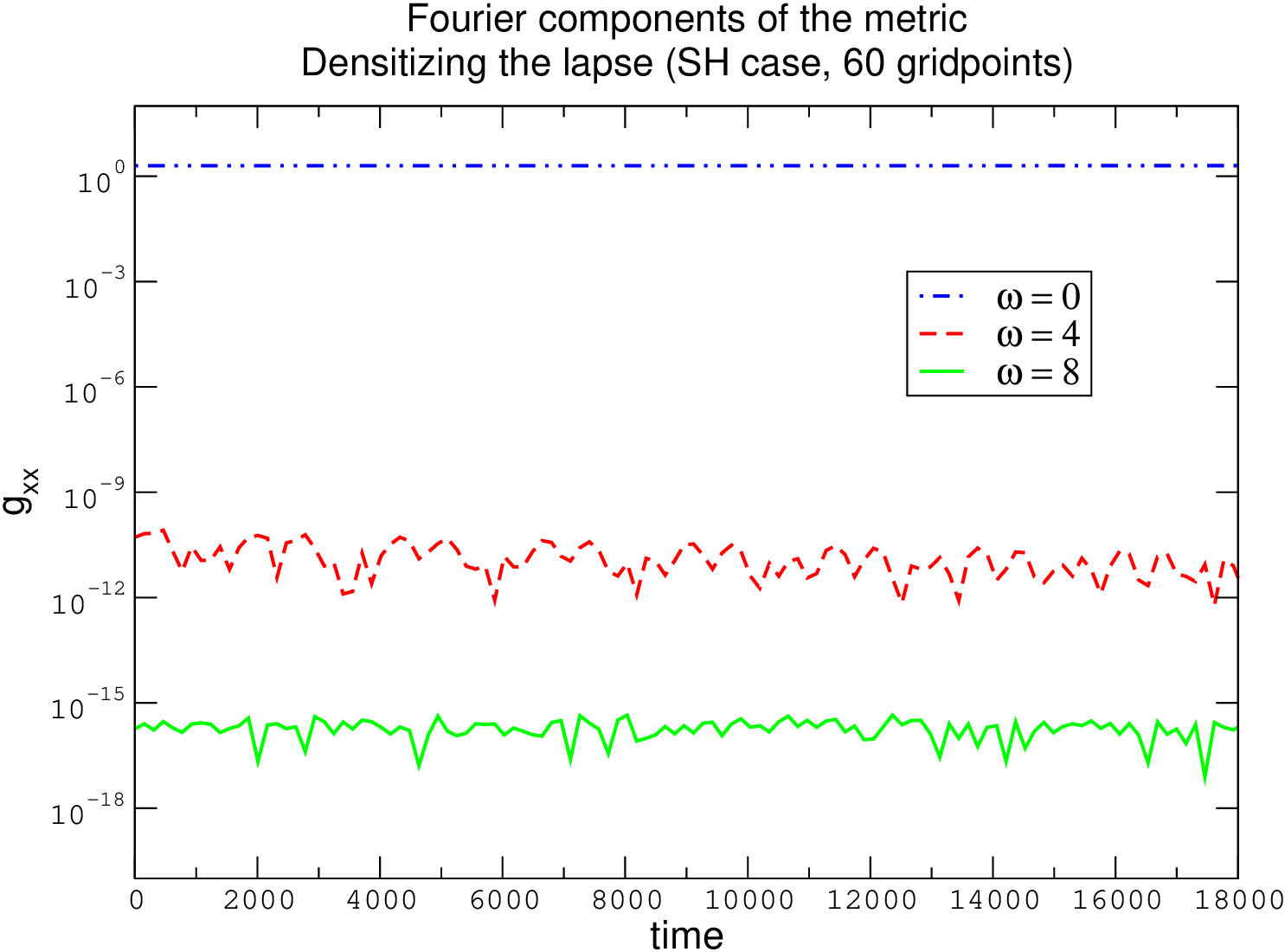}
\caption{Fourier components of the numerical metric for $\omega
=0,4,8$.  \label{speed_fourier}}
\end{center}
\end{figure}

The continuum equation (\ref{model_eq}) is neither dissipative nor dispersive.
Ideally, one would like the difference scheme to have the same
properties.  
For small values of $A$ the initial data essentially consists of only
one Fourier mode, $A\sin x$, which corresponds to $\beta = \pm1$.  If one uses,
say, $60$ gridpoints and a spatial domain that
extends from $-\pi $ to $\pi $, then one has that $\beta \Delta x = \pi / 30$, for which
Eq.(\ref{ampfactor}) predicts a 
damping of about $0.0032\%$ per crossing time.
Similarly, from Eq.(\ref{disc_speed}) one can see that for this
resolution and this initial data one should have a
wave loss (i.e. the numerical and exact solution have a phase
difference of $2\pi$) at 
approximately $t=4,600$, which corresponds to $750$ crossing times.
\begin{figure}[ht] 
\begin{center} 
\includegraphics*[height=6cm]{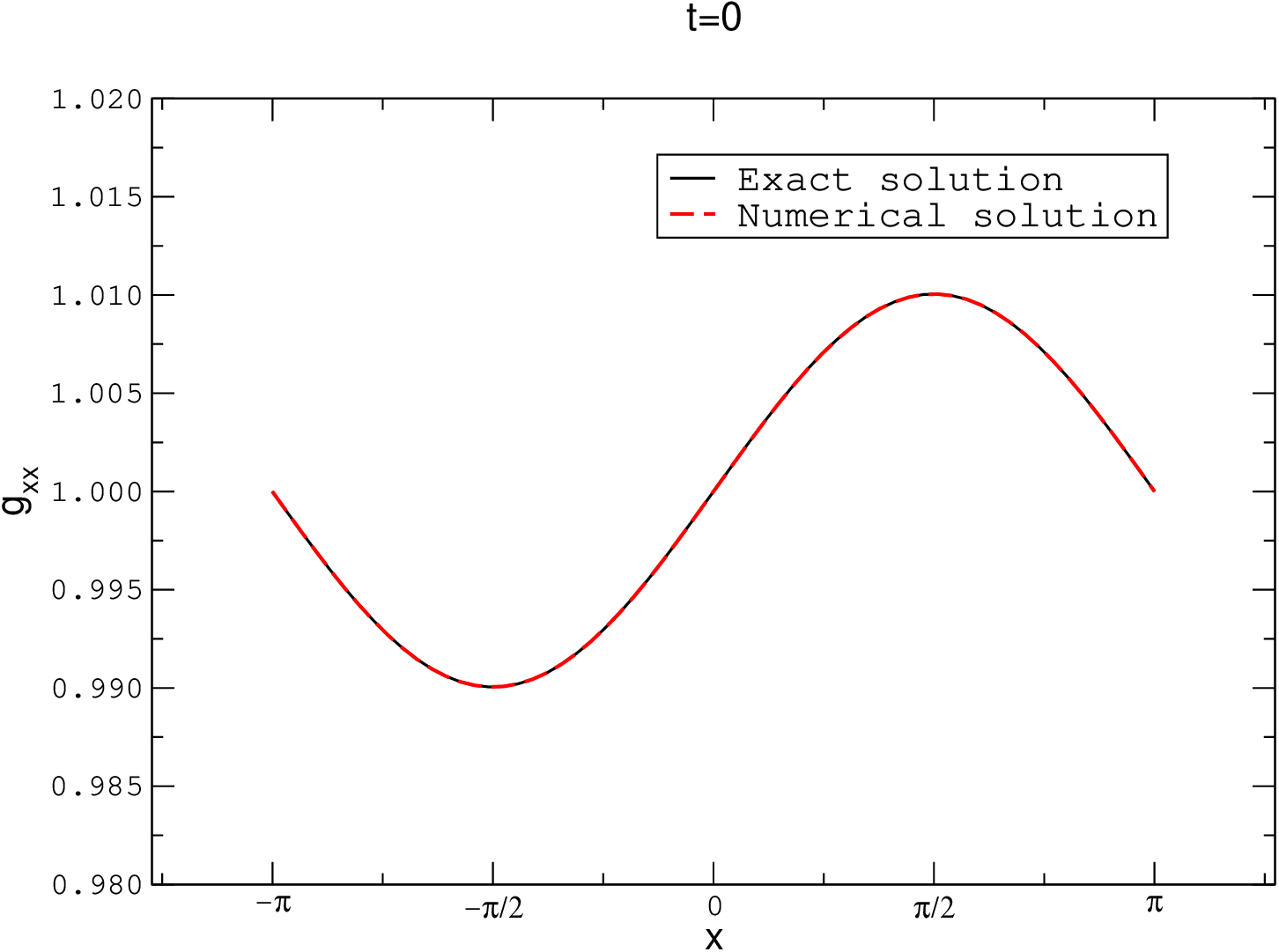}
\includegraphics*[height=6cm]{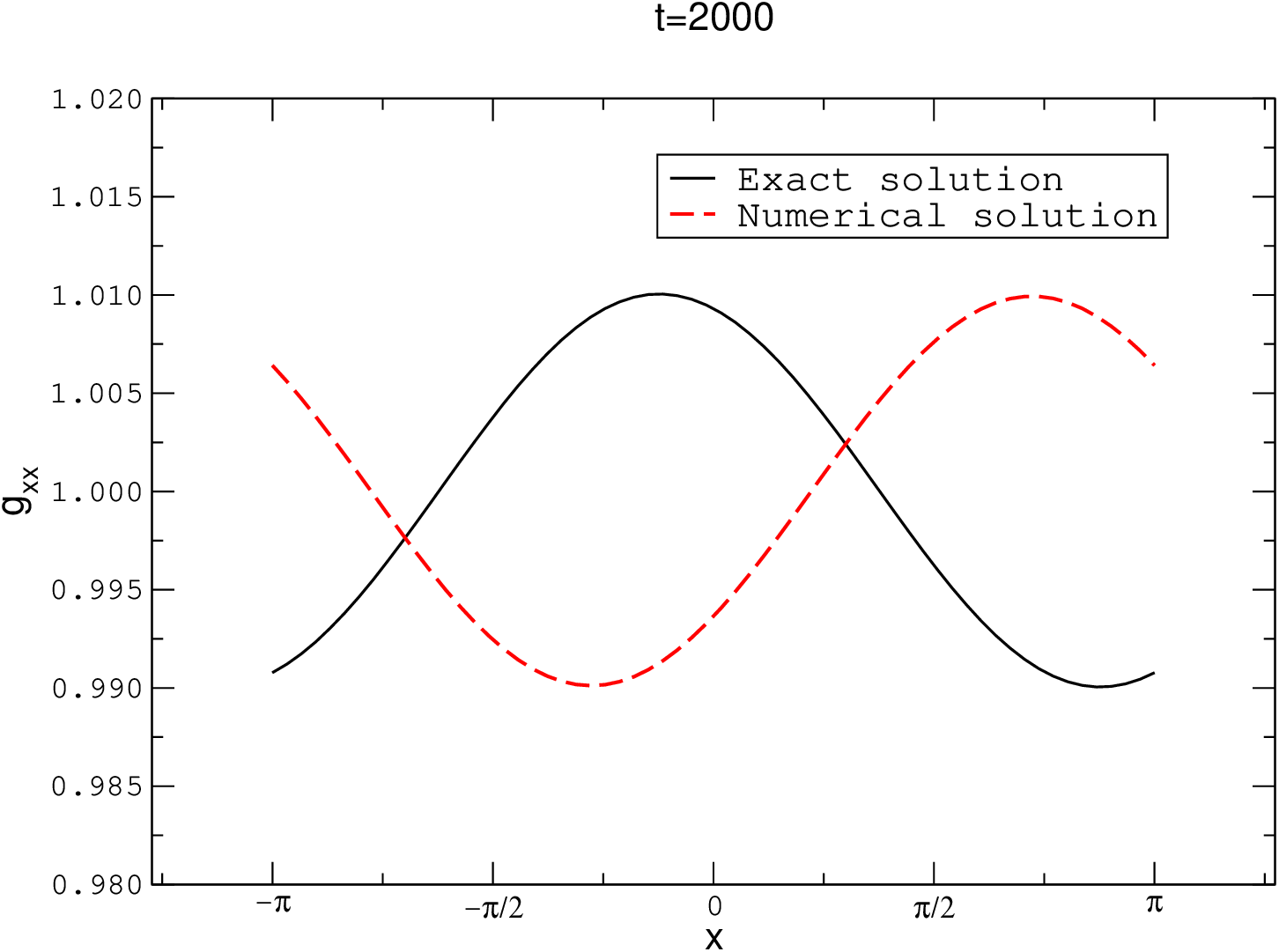}
\includegraphics*[height=6cm]{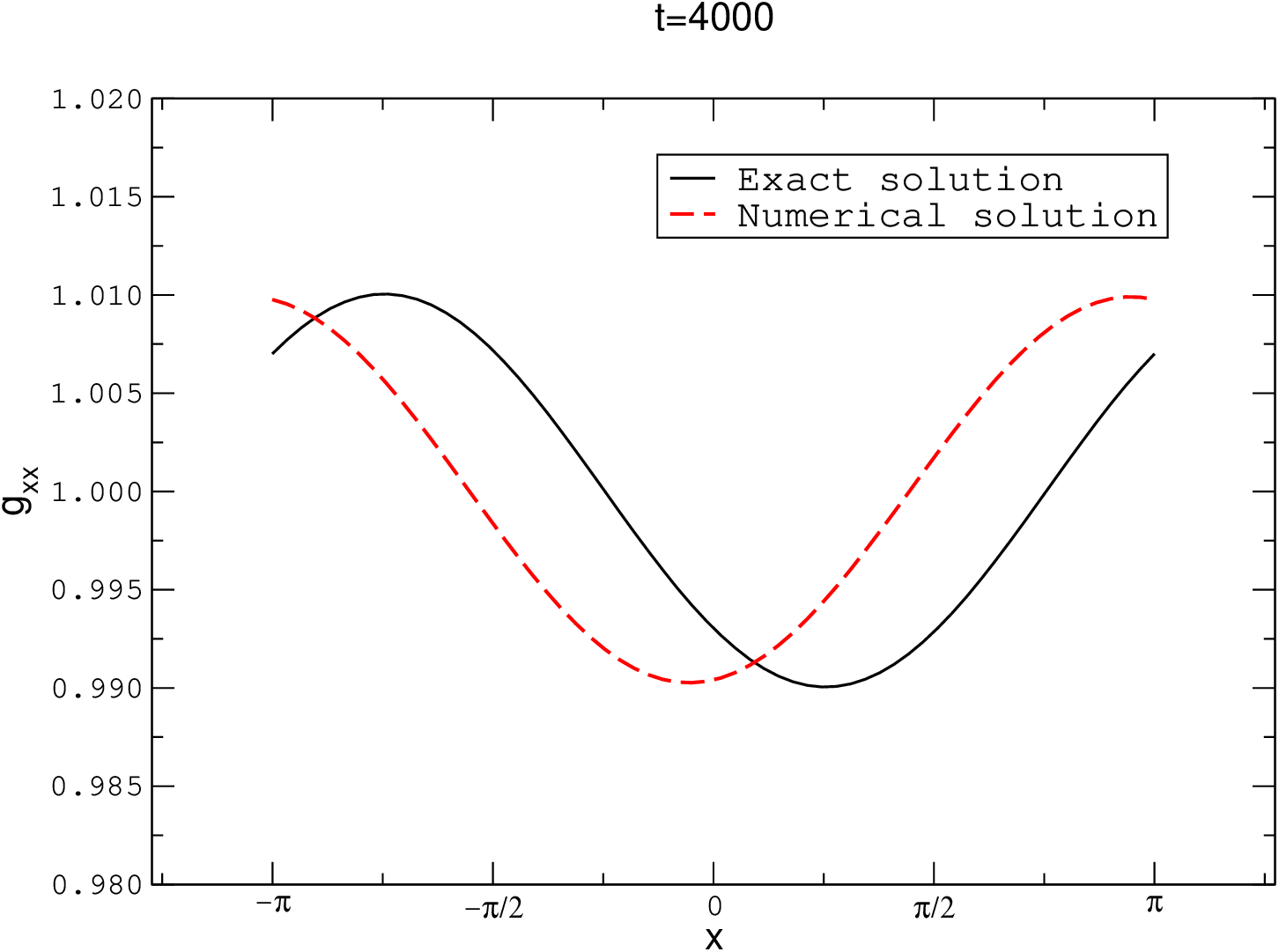}
\includegraphics*[height=6cm]{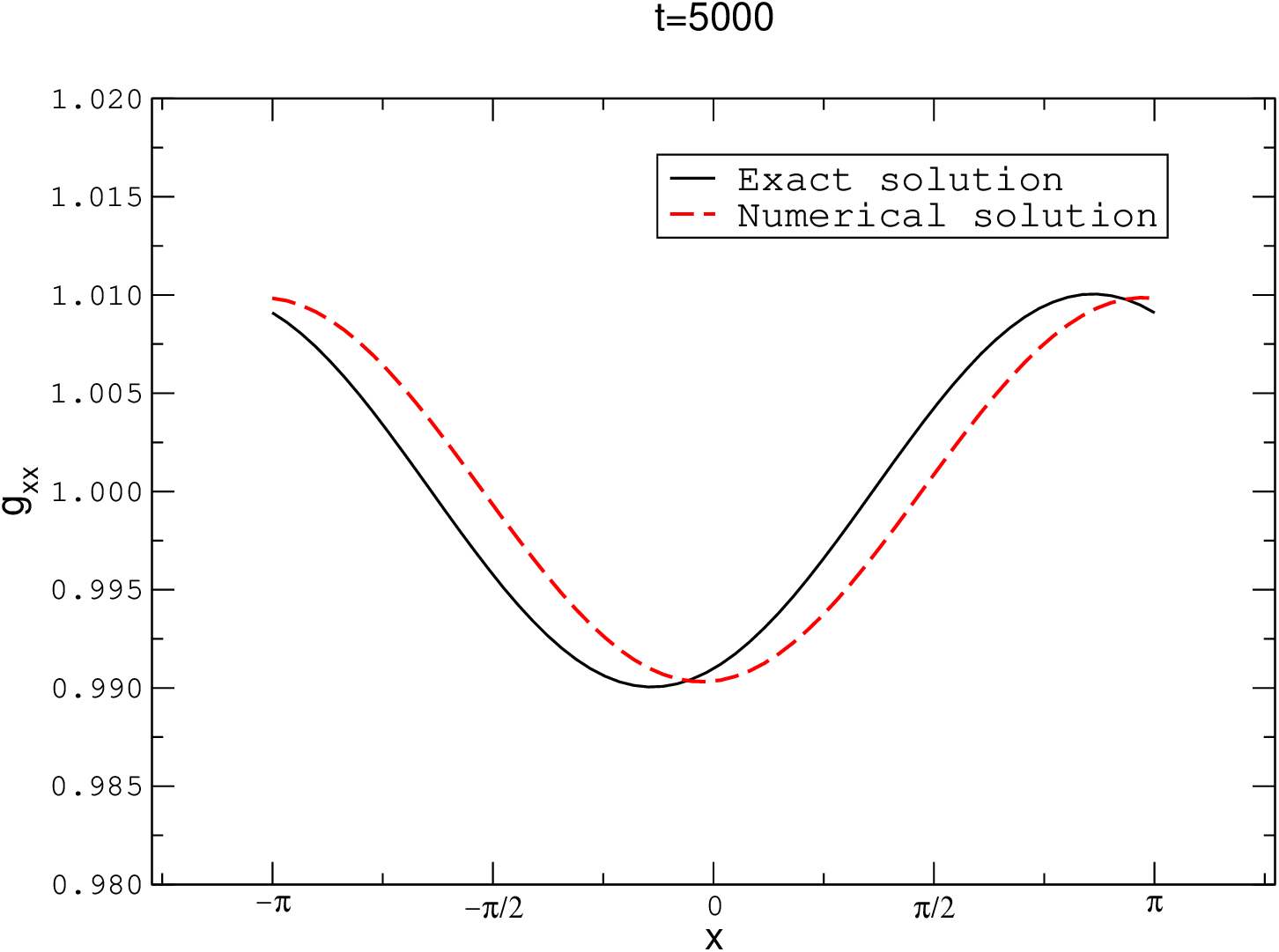}
\caption{Exact and numerical solutions for the SH case, at different times, showing
the speed difference. This run was done with $60$ grid points,
$\lambda =1/2$, $\tilde{\sigma} = 0.02$ and a domain of length $2\pi$. With
this resolution, the code `loses' one wave after, roughly, $750$
crossing times. \label{speed}} 
\end{center}
\end{figure}
Figure \ref{speed} show both the numerical and exact solution for the
$g_{xx}$ component of the metric at different times.  The predicted
speed difference is there and is in very good agreement with the
analytical calculation based on the model equation. Also,  the profile
of the numerical solution maintains its shape for a very long time.
Figure \ref{errors_sh_dens_gK} shows the $L_2$
norm of the errors for this component at the same resolution. The
errors come mostly from the speed difference, and the
maxima and minima correspond to a phase difference between the analytical and
numerical solution of odd and even multiples of $\pi$, respectively. Indeed, if one
computes the error in the metric due to a phase difference of $\pi$,
one gets
$$
\left(\int_{-\pi }^{\pi}\left(e^{\sin{\left(t+x\right)}/100}-
e^{\sin{\left(t+x+\pi \right)}/100}\right)^2dx \right)^{1/2}= 0.035
$$
which is in remarkable agreement as well with the maxima in figure
\ref{errors_sh_dens_gK}. 
\begin{figure}[ht] 
\begin{center} 
\includegraphics*[height=6cm]{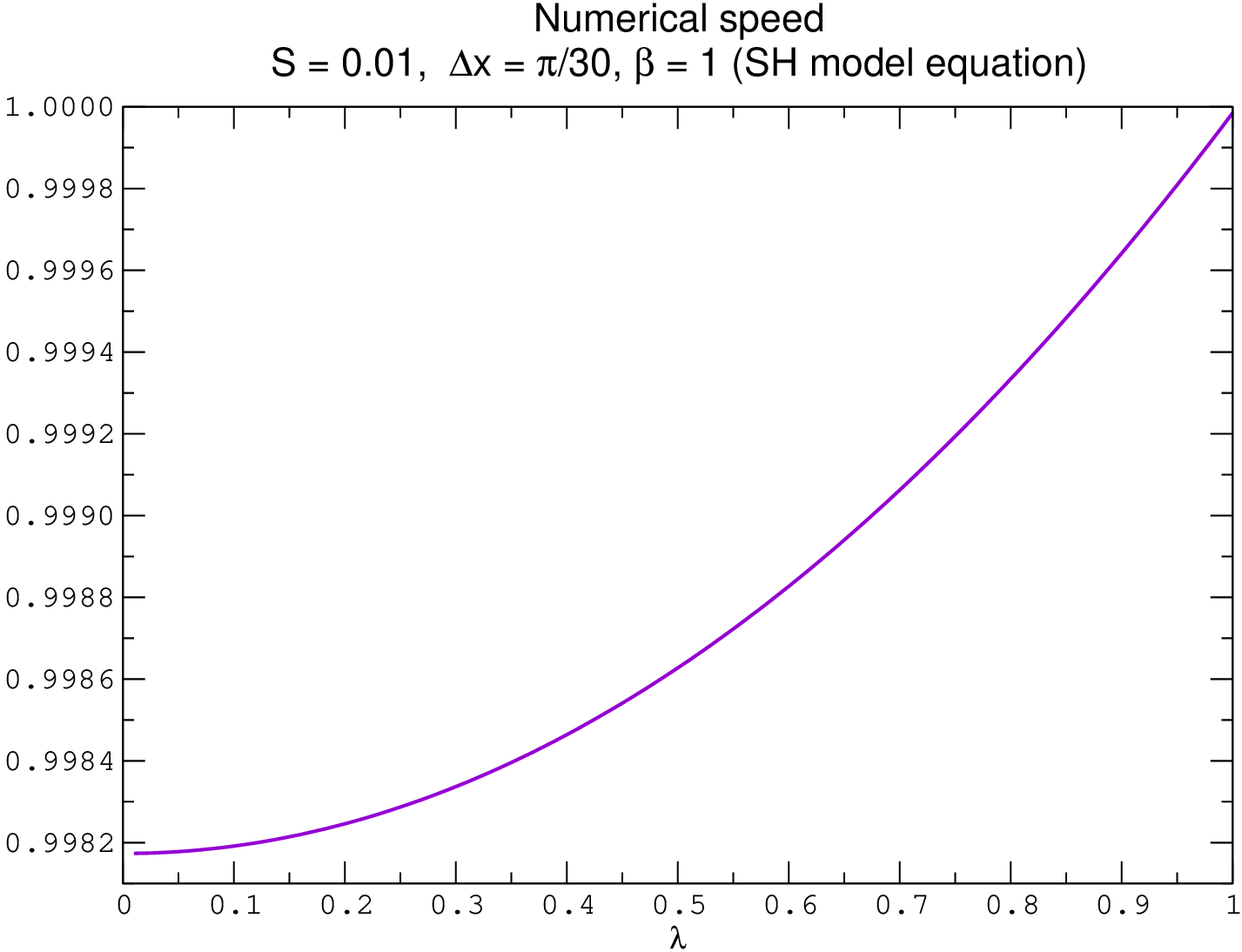}
\caption{Dependence of the numerical speed on the Courant factor.
Increasing the Courant 
factor, decreases the error in the propagation speed. \label{numspeed_R}}
\end{center}
\end{figure}
\begin{figure}[ht] 
\begin{center} 
\includegraphics*[height=6cm]{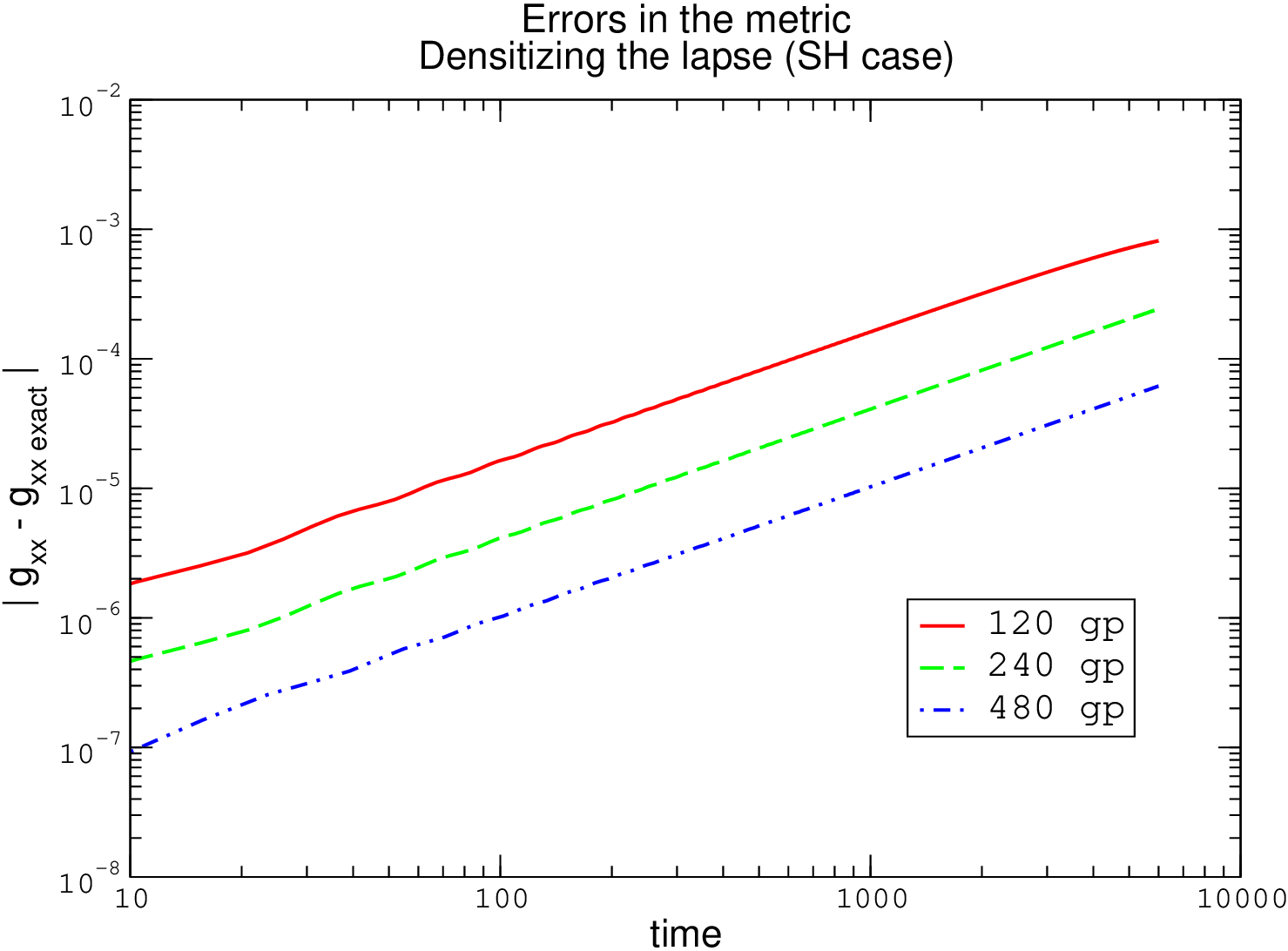}
\caption{$L_2$ norm of the errors for the metric. \label{errors_sh_dens_res}}
\end{center}
\end{figure}
This speed difference is a rather usual numerical
feature in solutions of 
finite difference schemes approximating hyperbolic equations. 
Usually this difference is
{\em smaller} for bigger Courant factors \cite{thomas}, as can be seen from figure
\ref{numspeed_R} for the particular discretization here used.   
One could try to push
this factor to its stability limit, or use a different scheme
that minimizes this effect.  But here the point that we want to
emphasize is that the errors in the SH simulation are second
order convergent ones, as shown in figure
\ref{errors_sh_dens_res}. That figure
shows the $L_2$ norm of the error in one component of the metric,
 $g_{xx}$, 
for resolutions ranging from $120$ to $480$ gridpoints, up to $1,000$
crossing times (in order to evolve up to $1,000$ crossing times
without loosing a wave, one has to use more than $60$ gridpoints).
One can see that the errors grow
linearly in time. One could
be tempted to think that this is due to the zero-frequency linear mode
predicted in Section \ref{kst}. This is not the case, in this
simulation this mode is not excited (though it could be, in a more
general evolution) and the error is caused by
the numerical speed difference, as discussed above. Indeed, by performing
a Fourier decomposition in space of the
numerical solution one explicitly sees that the amplitudes are
roughly constant in time (for non zero frequency components, this is exactly
what the linearized analysis at the continuum predicted). Figure \ref{speed_fourier}
shows some of these
amplitudes for this simulation (to be contrasted later with the WH and
CIP cases).
\begin{figure}[ht] 
\begin{center} 
\includegraphics*[height=6cm]{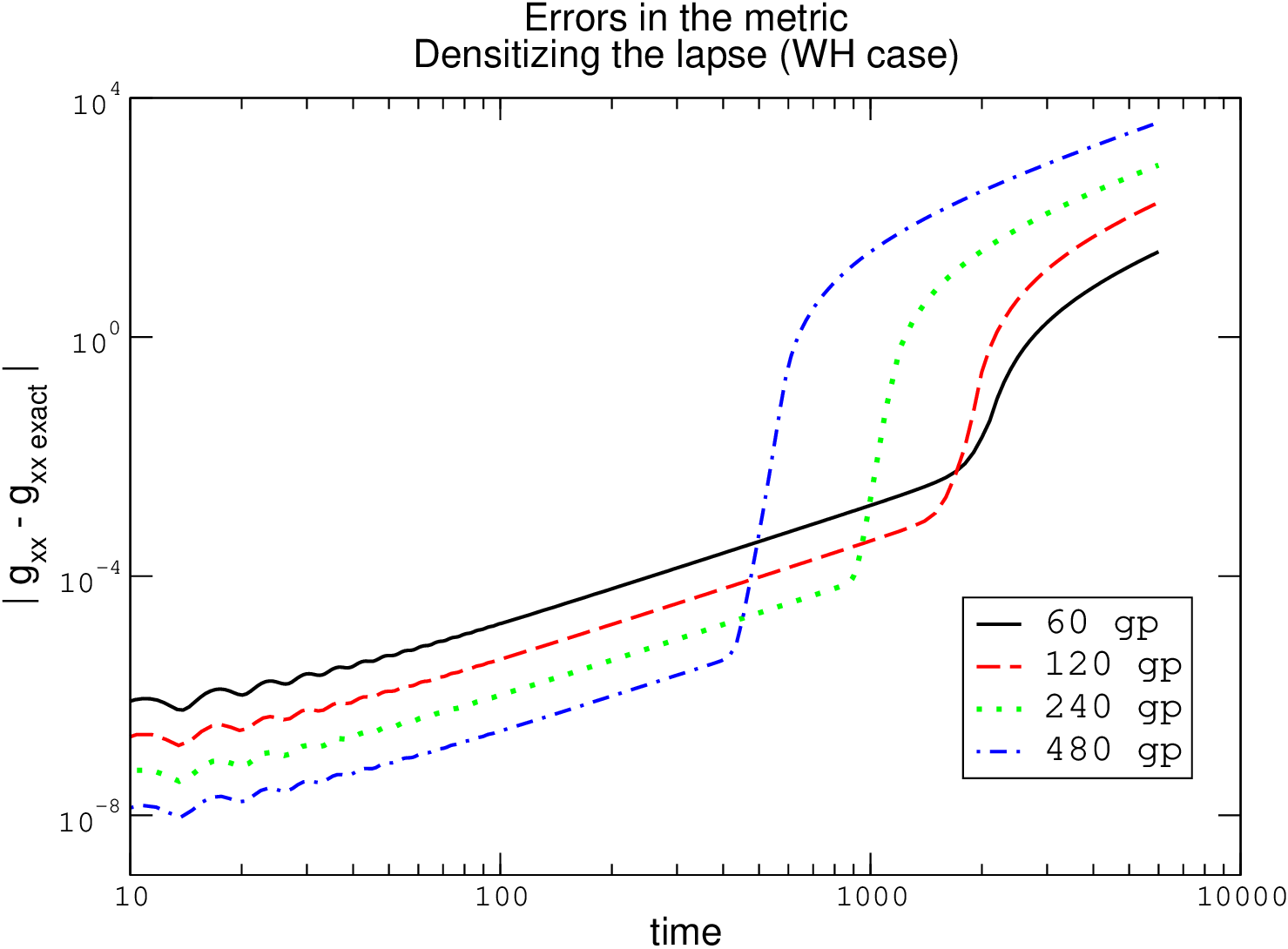}
\caption{$L_2$ norms of the errors for the metric. \label{errors_wh_dens}}
\end{center}
\end{figure}

\subsubsection{The WH case ($\sigma=0$)} 

Figure \ref{errors_wh_dens} shows plots of the errors associated with evolutions
performed with the same
initial data, dissipation and Courant factor as above, except that
now we densitize the lapse according to $\sigma =0$ (``exact
lapse''). 
\begin{figure}[ht] 
\begin{center} 
\includegraphics*[height=6cm]{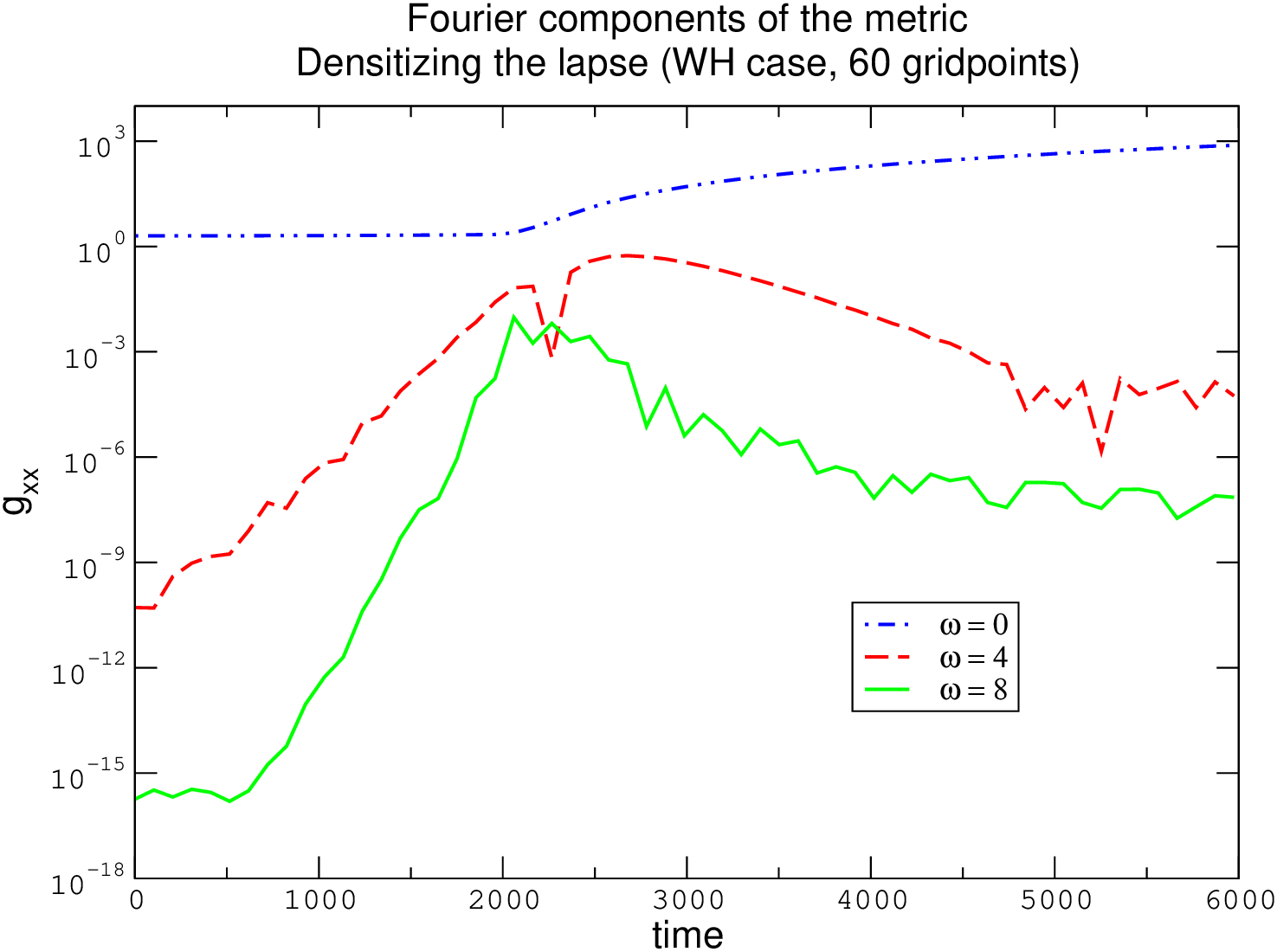}
\caption{Fourier components of the numerical metric for $\omega
=0,4,8$.  Some of the components grow exponentially 
from the very beginning.\label{fourier_wh_dens_60gp}}
\end{center}
\end{figure}
For two fixed resolutions the coarsest one gives smaller errors after
a while, and the time at which 
this occurs decreases as one increases resolution.  This indicates
that the difference scheme {\em is not convergent}.  Note that one
could be easily misleaded to think that the code is convergent 
if one did not evolve the system for long enough time, or without high 
enough resolution. For example, if one performs two runs, with $120$
and $240$ gridpoints, one has to evolve until, roughly, $150$ crossing
times, in order to notice the lack of convergence.  To put these numbers 
in context, suppose one had a
similar situation in a 3D black hole  evolution. To give some
usual numbers, suppose the 
singularity is excised, with the inner boundary at, say,
$r=M$, and 
the outer boundary is at $20M$ 
(which is quite a modest value if one wants to extract
waveforms). In this case $120$ and $240$ gridpoints correspond to grid
spacings of, approximately, $M/5$ and $M/10$, respectively (usual
values as well in some simulations). If one had
to evolve up to $150$ crossing times in order to notice the lack of
convergence, that would correspond to $t=3,000M$, which is several times 
more than what present 3D evolutions last. 
Of course, the situation presented in this
simple example need not appear in exactly the same way in an evolution
of a different spacetime, or with a different discretization; in fact,
in the next subsection we show an example where the instability becomes obvious sooner. Also,
there are some ways of noticing in advance
that the code is not converging. Namely, it seems that the numerical solution has
the expected power law growth that the continuum linearized analysis predicts
until all of a sudden an exponential growth appears. But if one
looks at the Fourier components of the numerical solution, one finds
that there are non-zero components growing exponentially from the very
beginning, starting at the order of truncation error (see figure \ref{fourier_wh_dens_60gp}).

One might expect that, since for WH systems the frequency-dependent
growth at the continuum is a power
law one, it is possible to get convergence by adjusting the 
dissipation. In \cite{convergence} we show that even though certain
amount of dissipation might help, the code is never 
convergent and, 
indeed, adding too much dissipation violates the von Neumann
condition, which leads to a much more severe numerical
instability. We have systematically done
numerical experiments changing the value of $\tilde{\sigma}$ without being
able to stabilize the simulations (more details are given
below), verifying, thus, the discrete predictions.

\subsubsection{The CIP case ($\sigma =-1/2$)} 

\begin{figure}[ht] 
\begin{center} 
\includegraphics*[height=6cm]{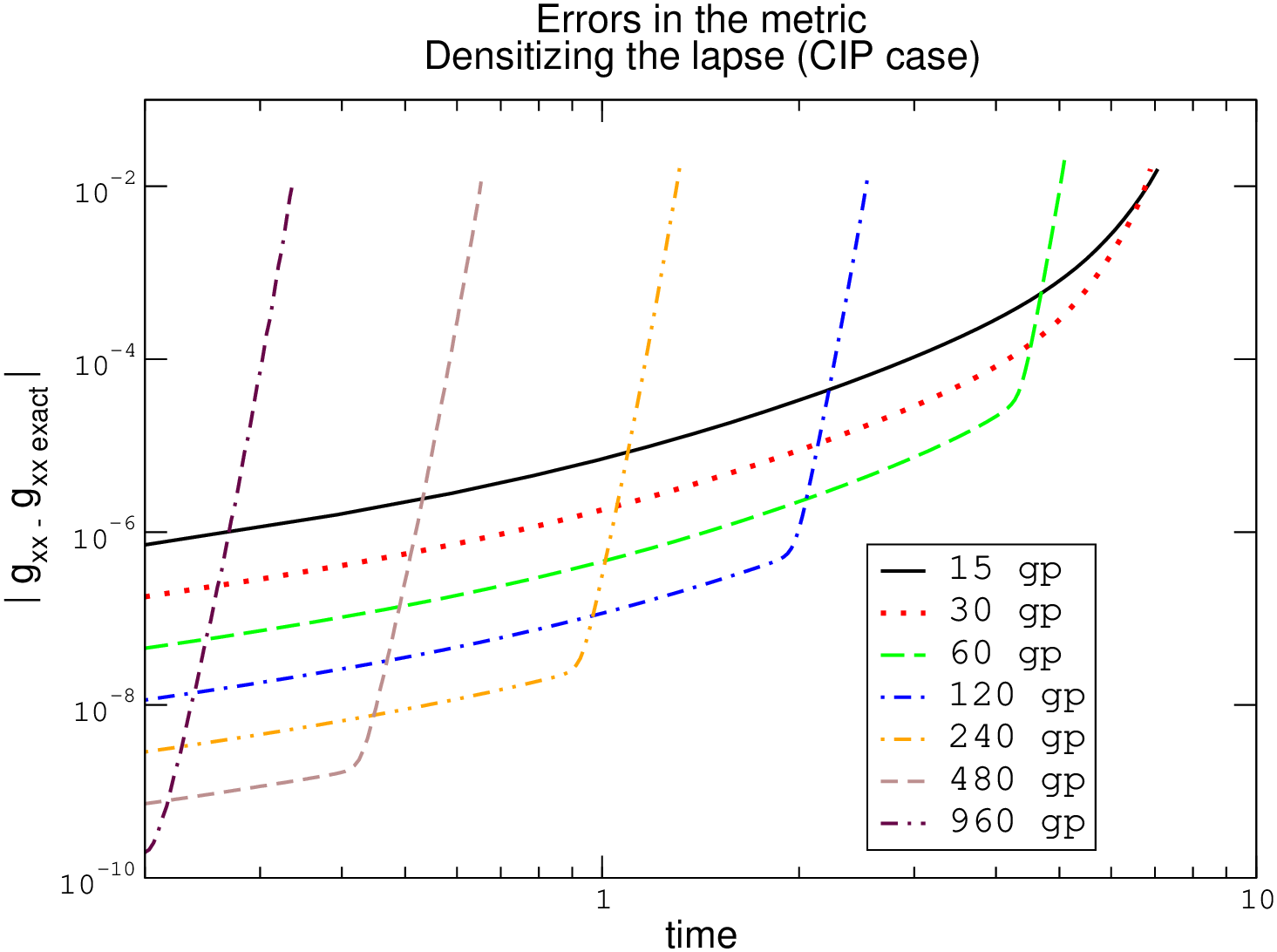} 
\caption{$L_2$ norm of the errors for the metric.
\label{errors_ip_dens}}
\end{center}
\end{figure}

Figure \ref{errors_ip_dens} shows the error in the metric, for
different resolutions. As in the WH case, 
the errors originate mostly from the non-zero frequencies (i.e.~the ones that
typically grow in an unstable numerical scheme). But now they
grow more than 10 orders of magnitude in much less than one crossing time and
it is quite obvious that the code is not converging. This is so
because in the CIP case the instability grows exponentially with  the
number of gridpoints. 
\begin{figure}[ht] 
\begin{center} 
\includegraphics*[height=6cm]{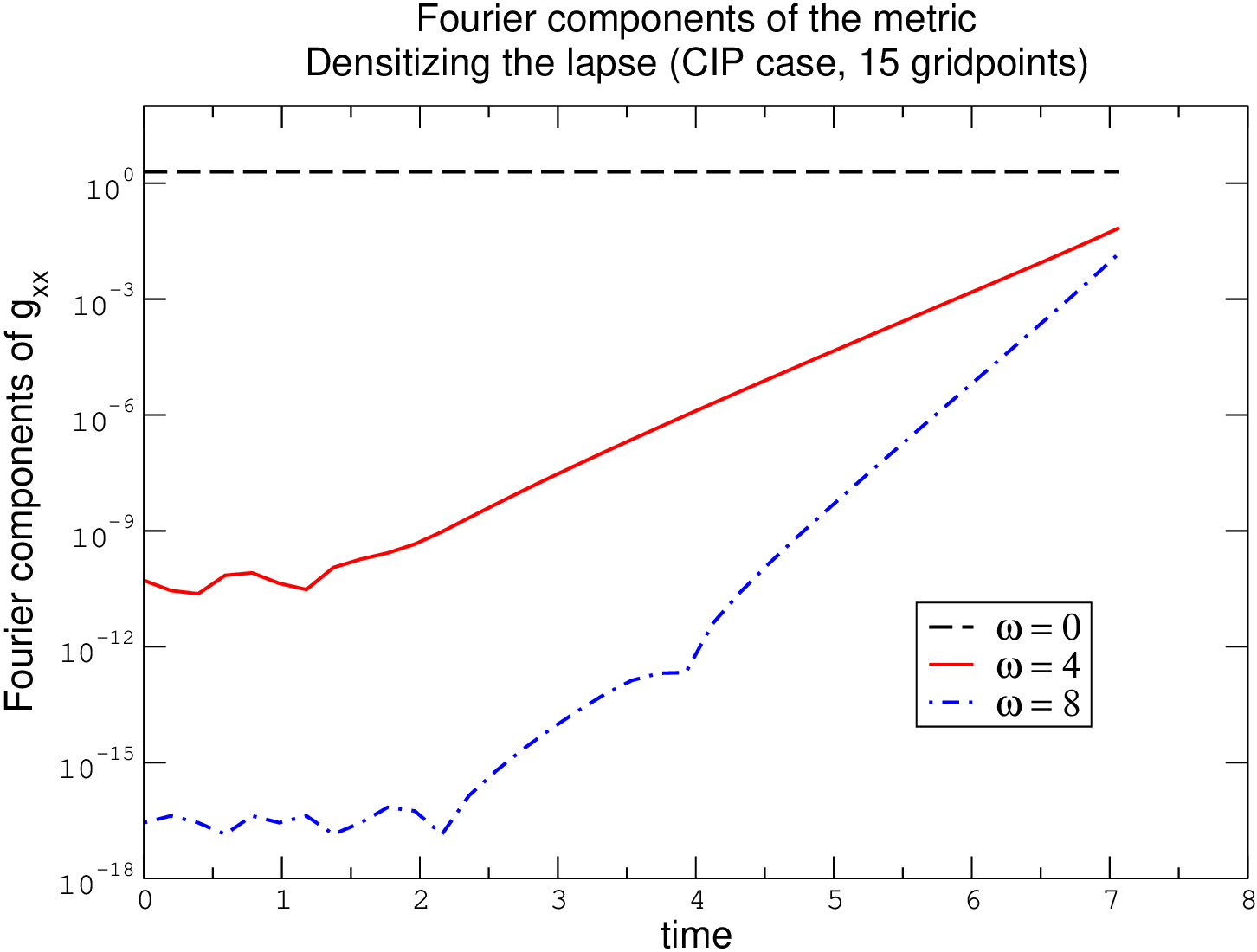}
\includegraphics*[height=6cm]{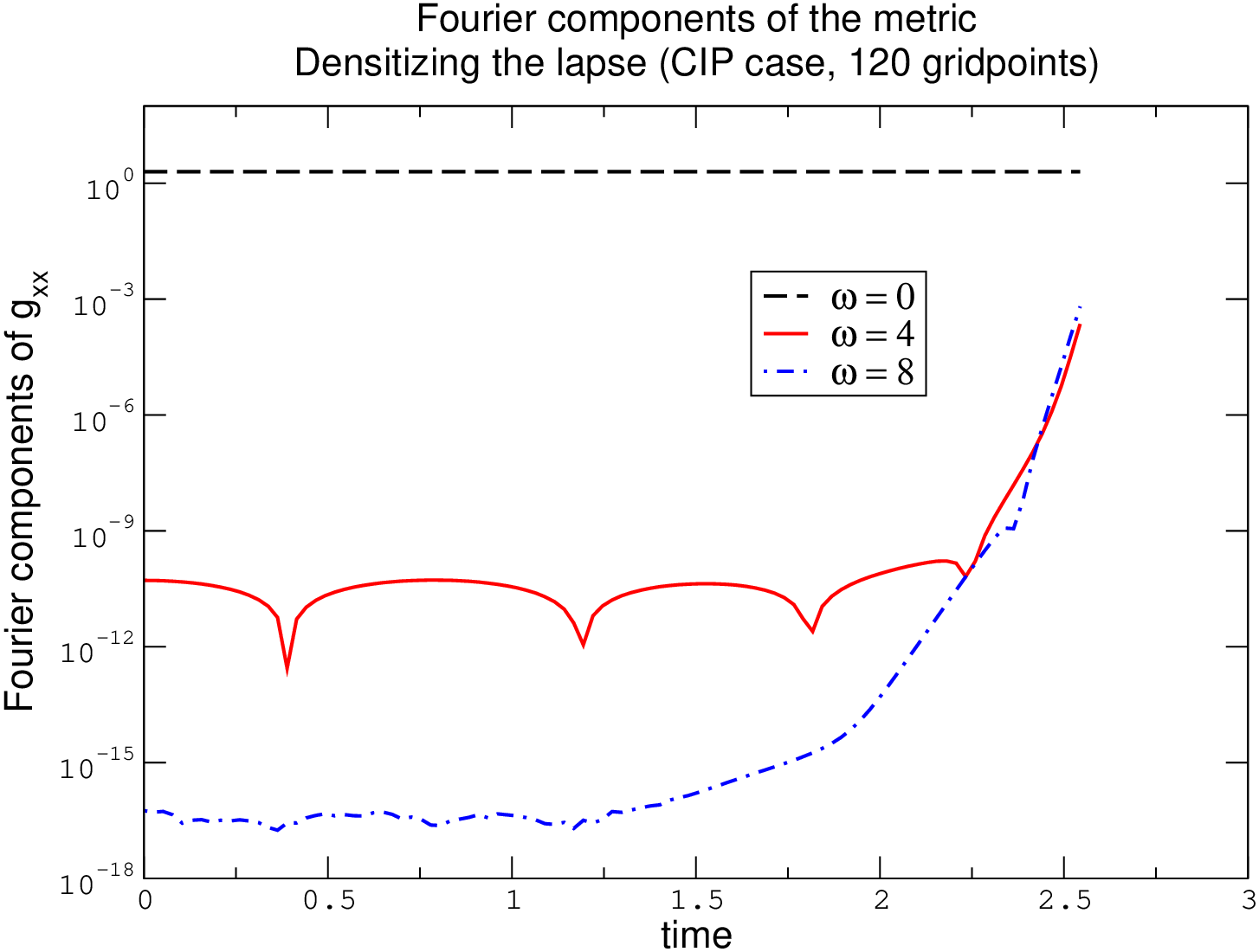}
\includegraphics*[height=6cm]{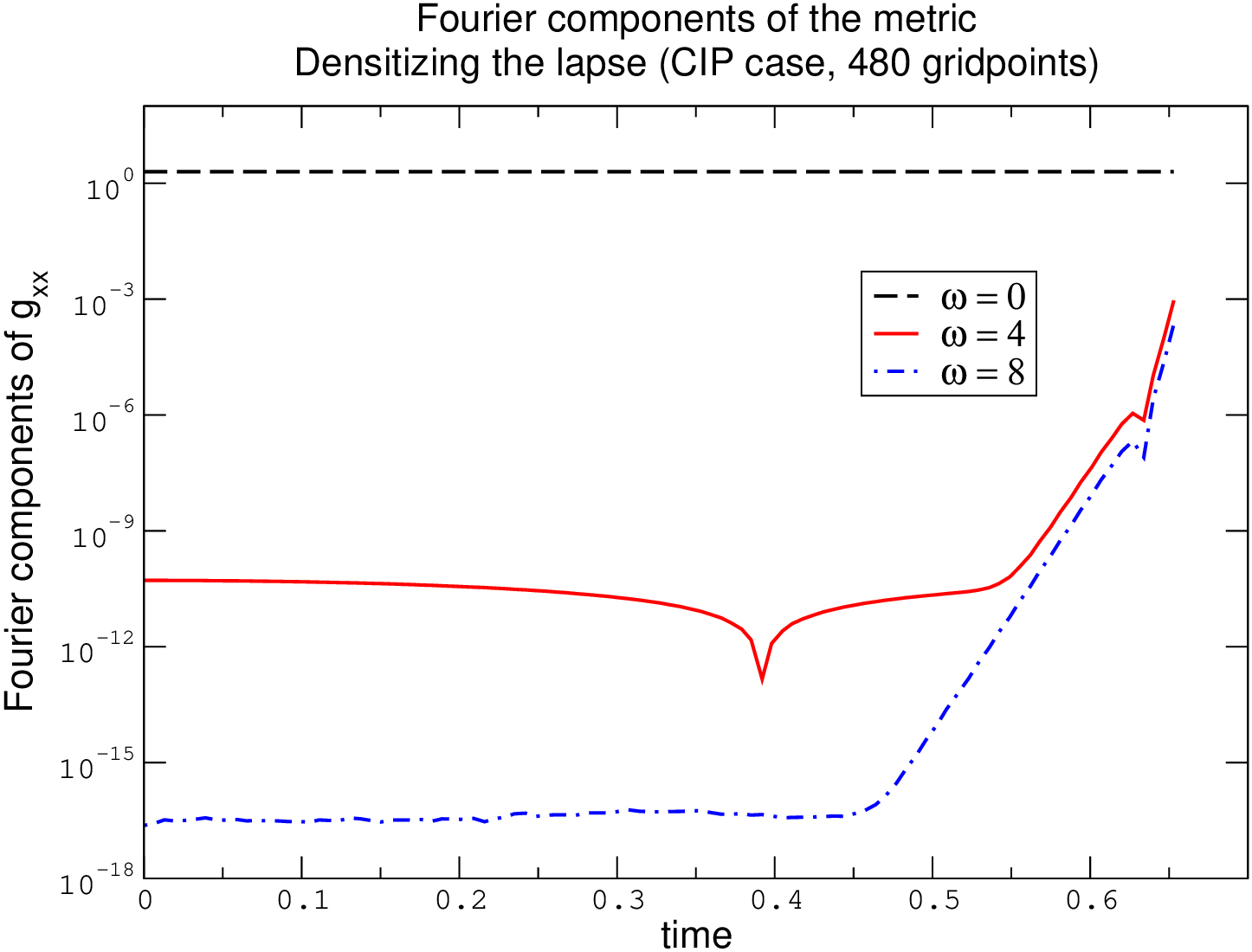}
\caption{The rate of growth of the Fourier components increases as one increases the resolution.
\label{fourier_ip_dens}}
\end{center}
\end{figure}
This can be seen performing a discrete analysis for the 
single ill posed equation in 1D, $v_t=iv_x$.
One gets that the symbol $\rho(\beta \Delta x)$ is
real and cannot be bounded by $1$ in magnitude, making the difference
scheme unstable (independently of resolution). 
\begin{figure}[ht] 
\begin{center} 
\includegraphics*[height=6cm]{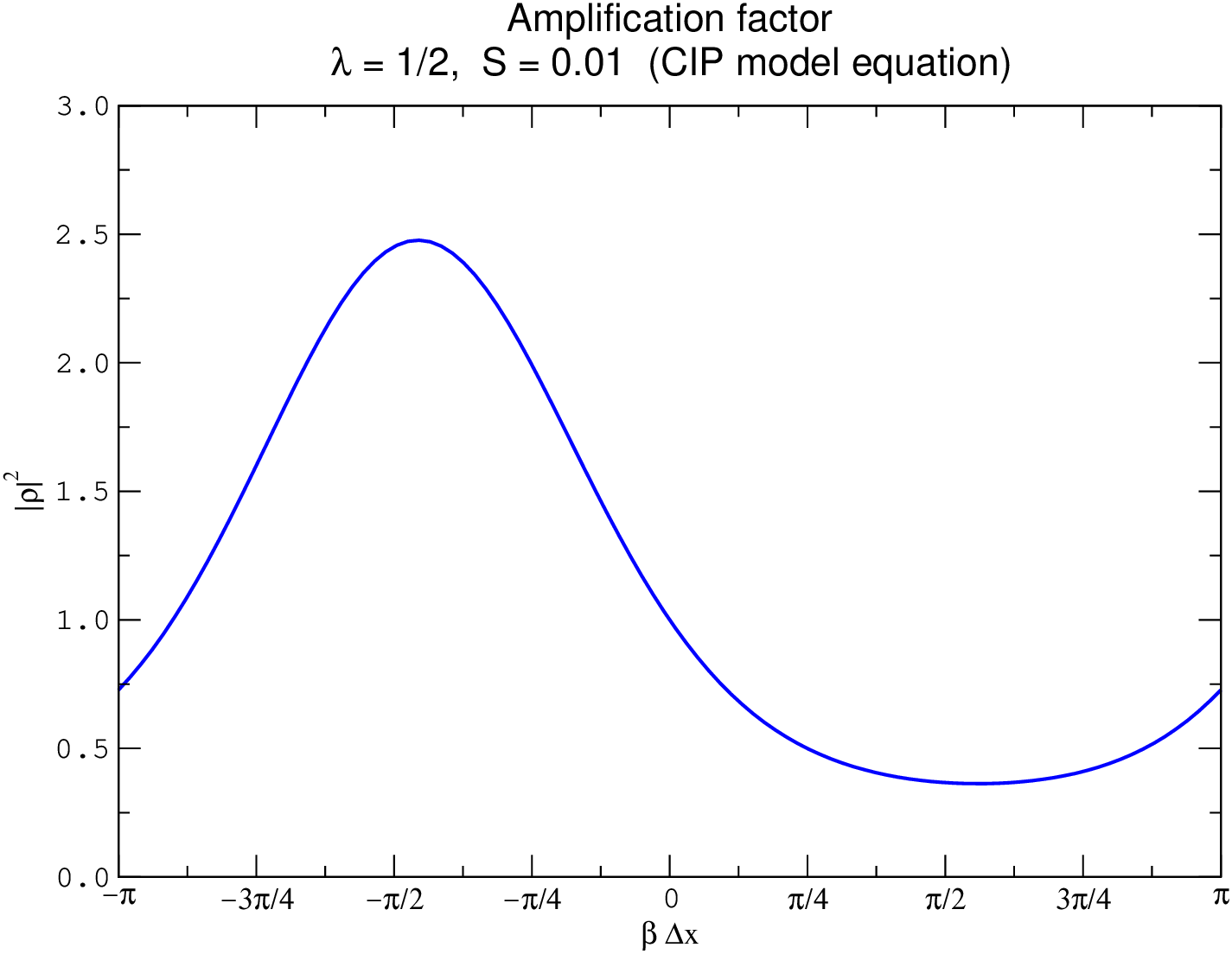}
\caption{Amplification factor associated with the difference scheme
(\ref{RK}) approximating the ill posed equation $v_t = i v_x$.  \label{ampl_IP}}
\end{center}
\end{figure}
If one changed to
characteristic variables exactly this equation would appear in 1D as a subset of
the system that we are considering, so this model equation is, in the
linear approximation,
part of the system evolved in this subsection. The amplification
factor for this  model equation is plotted in figure \ref{ampl_IP}.

\subsection{Adding the constraints}

Now we fix all parameters except the one  that
corresponds to the addition of the Hamiltonian constraint,
and choose three values that will render SH, WH and CIP systems.
The results are very similar to those of the previous section,
so we will present them in a less detailed way,
except that now we will discuss the role of explicit
dissipation. 

\subsubsection{The SH case}
\begin{figure}[ht] 
\begin{center} 
\includegraphics*[height=6cm]{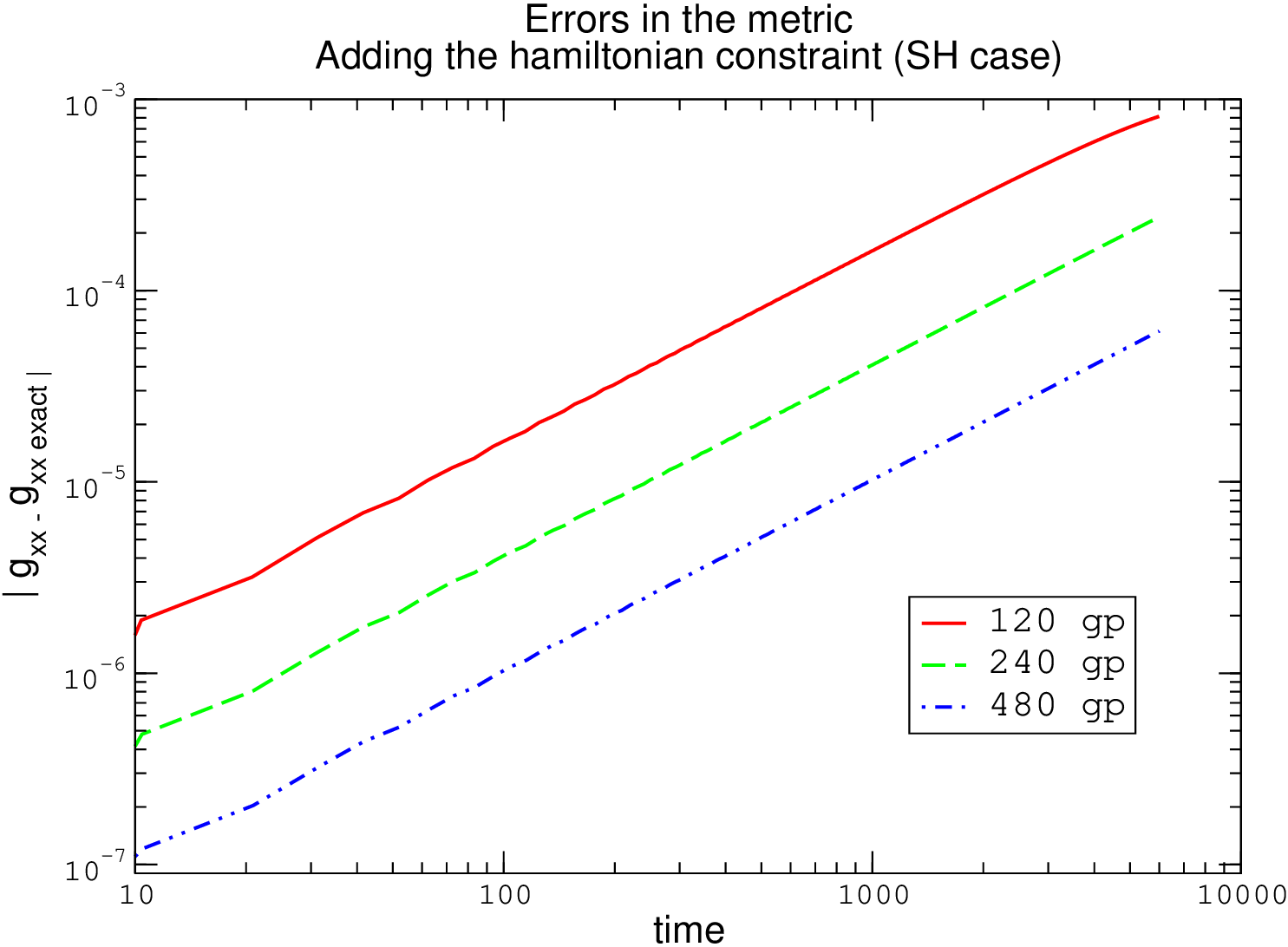} 
\caption{$L_2$ norm of the errors for the metric. \label{errors_sh_ham}}
\end{center}
\end{figure}

We start by considering the one-parametric KST subfamily of strongly hyperbolic
formulations of Eq.(2.39) of KST:
\begin{equation}
\sigma = \frac{1}{2}\, , \;\;\;
\zeta = -\frac{1}{9}(23+20\chi), \;\;\;
\gamma = -\frac{7}{6}\, \;\;\;
\eta = \frac{6}{5}\, \label{par_sh}.
\end{equation}
This family has, for any value of $\chi $, characteristic speeds $0$
or $\pm 1$ (see Section \ref{kst}), for the simulations here shown we chose $\chi =1$ (and
$\lambda =1/2 $, $\tilde {\sigma }=0.02$, as in the SH simulations previously discussed).
Figure \ref{errors_sh_ham} shows the analog of figure
\ref{errors_sh_dens_res}. As before, we do have convergence.  The Fourier
components of the solution remain constant in time, and the error
comes mostly from the numerical speed difference. 

\subsubsection{The WH case}
\begin{figure}[ht] 
\begin{center} 
\includegraphics*[height=6cm]{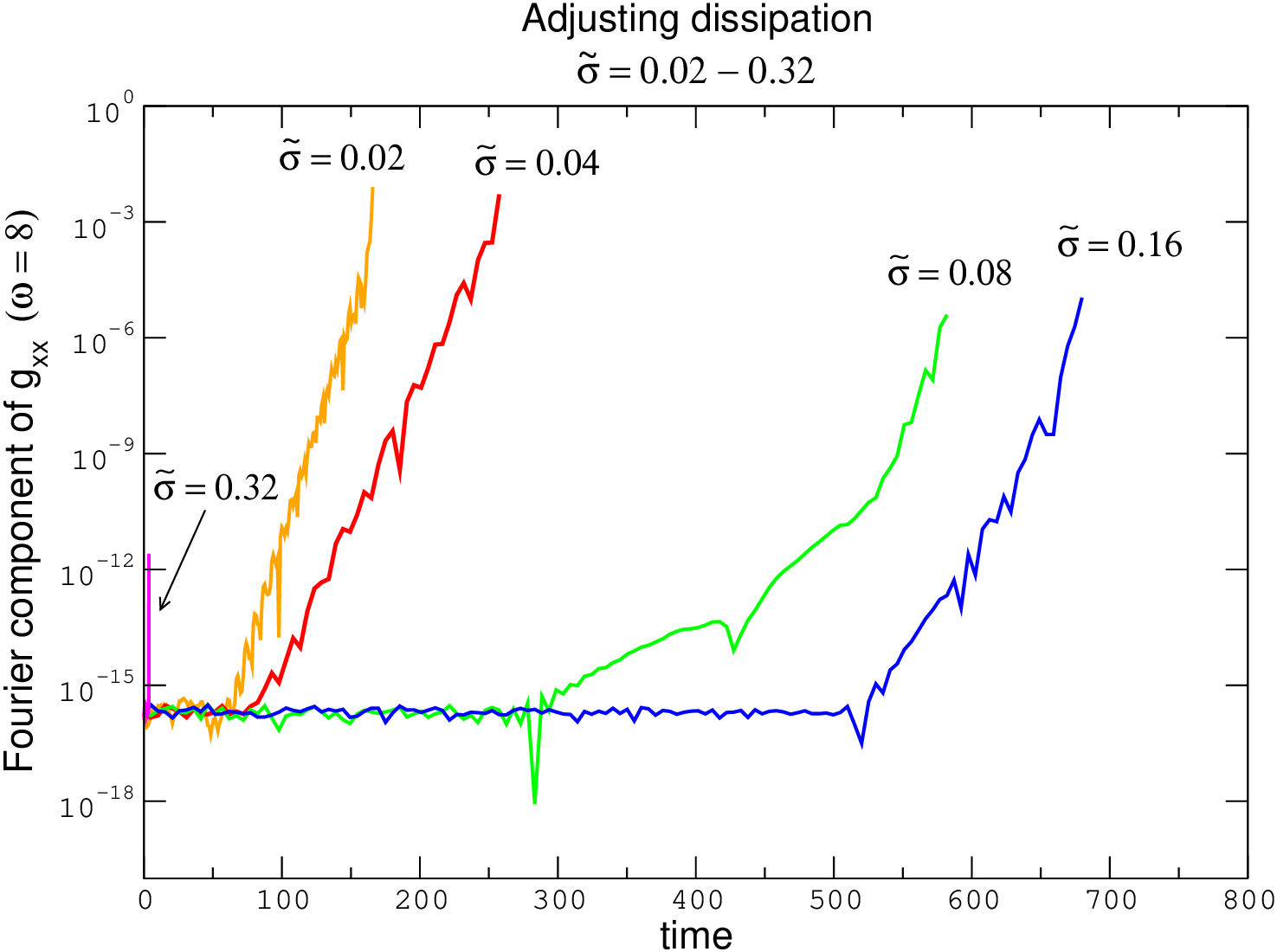} 
\includegraphics*[height=6cm]{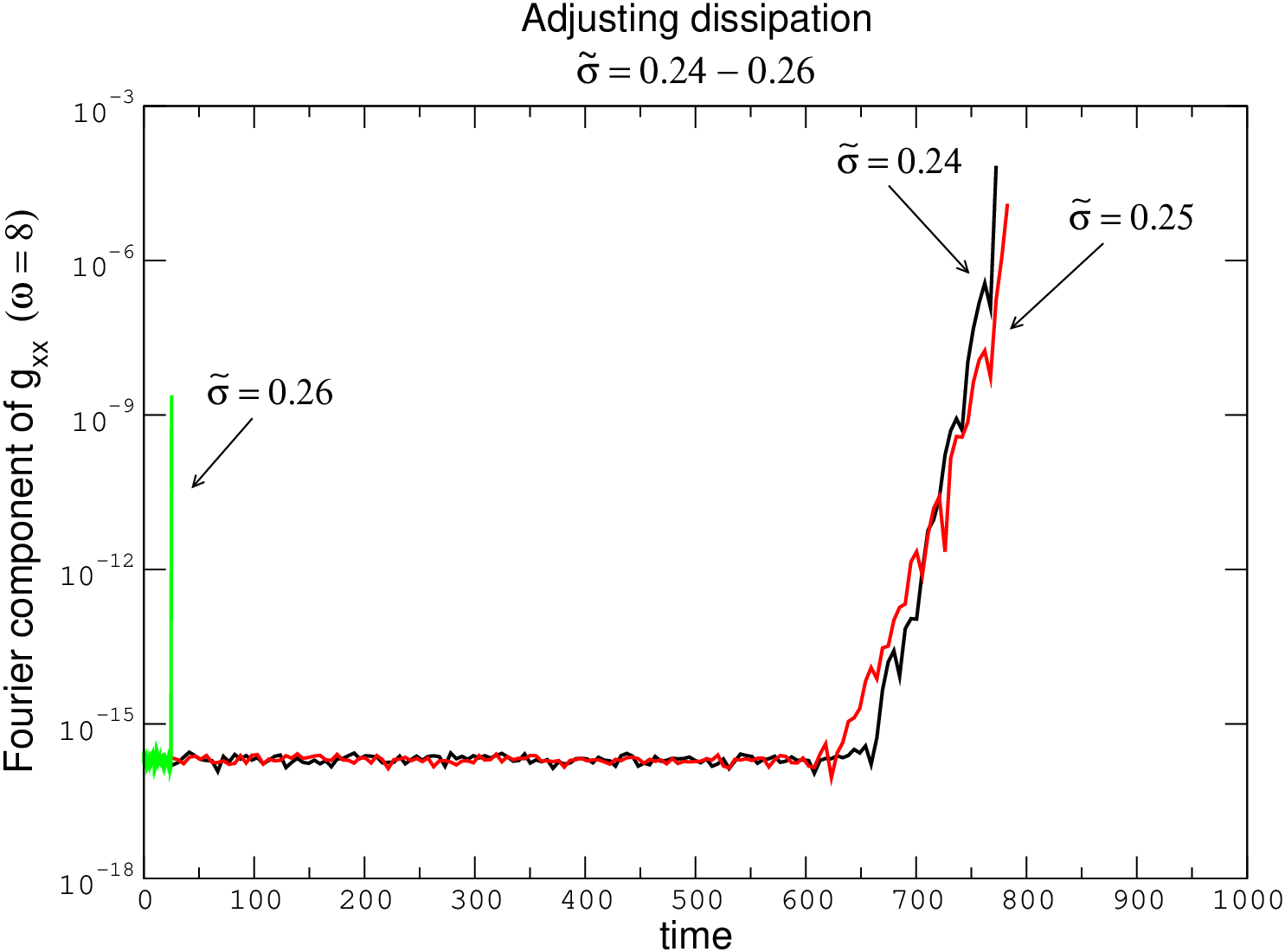} 
\includegraphics*[height=6cm]{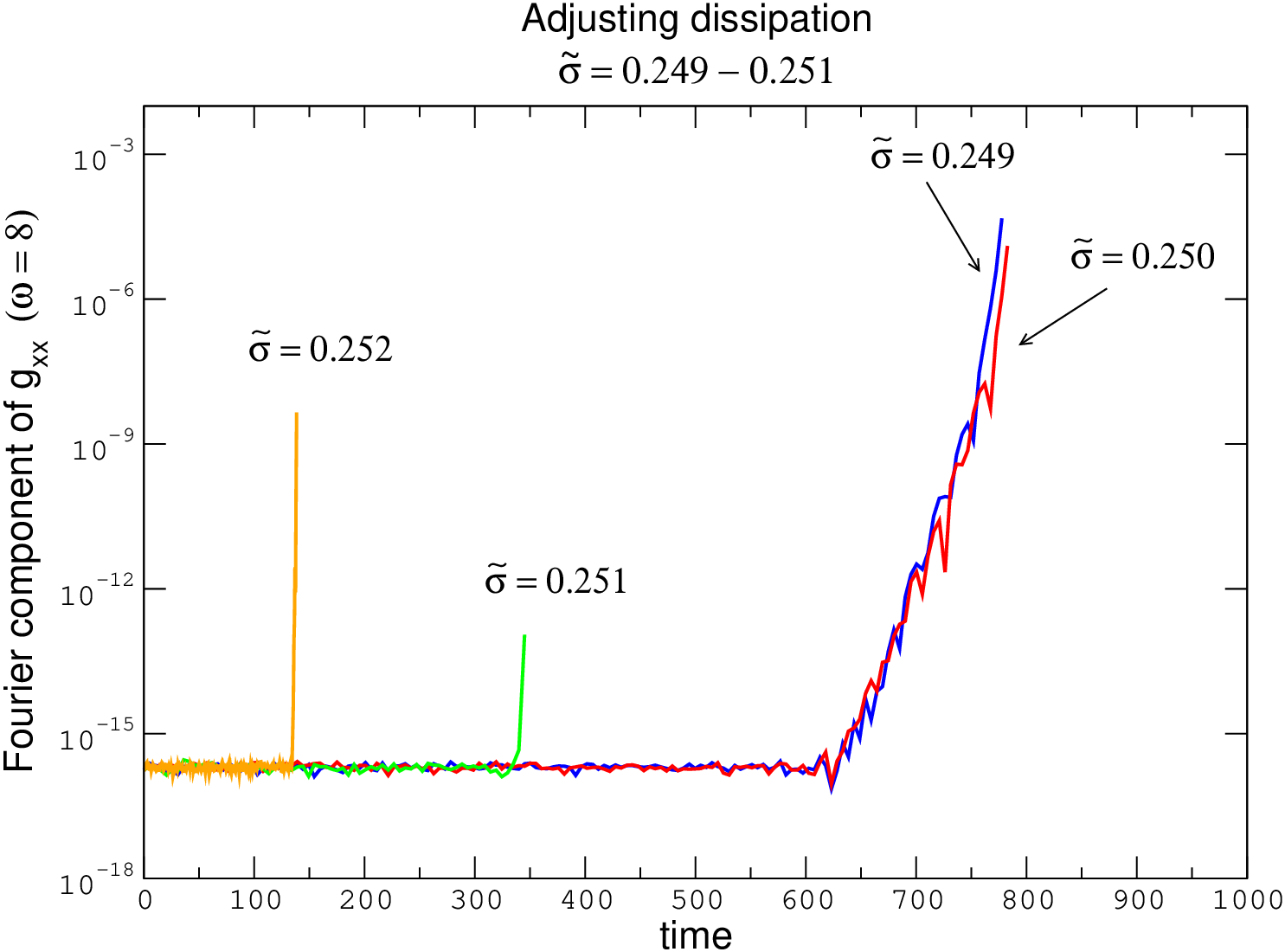}
\caption{We were not able to make the code stable by fine tuning the dissipation parameter.  \label{dissipation1}}
\end{center}
\end{figure}

Here we also use the parameters given by (\ref{par_sh}) with $\chi =1$, except that 
now we choose  
$\gamma = -32/21$. With this choice, $\lambda _1= \lambda _3 =1$, but
$\lambda _2=0$; therefore, as summarized in Section \ref{kst}, the system is WH. 
\begin{figure}[ht] 
\begin{center} 
\includegraphics*[height=6cm]{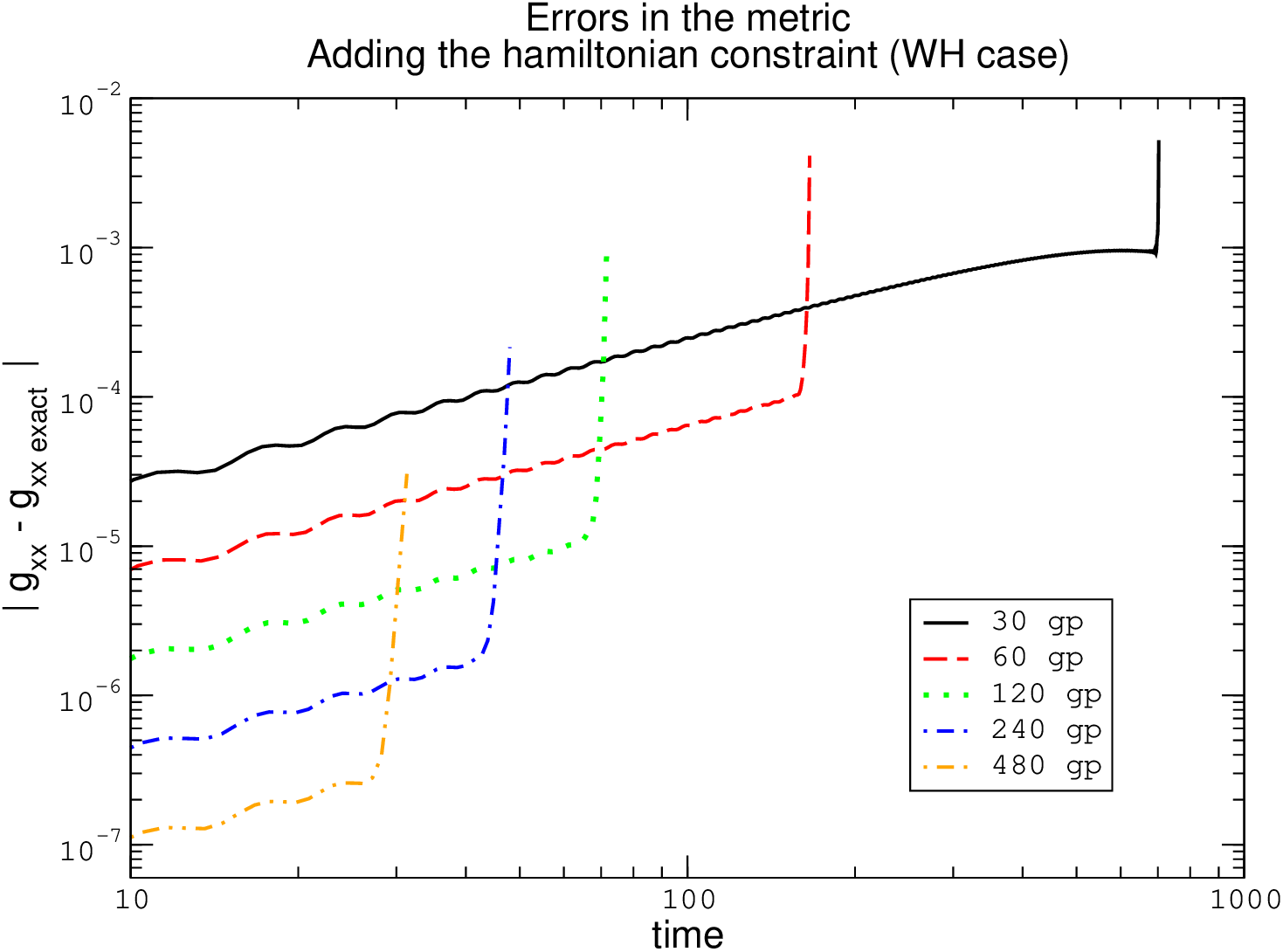} 
\caption{$L_2$ norm of the errors for the metric.  The simulation is
stopped once the determinant of the spatial
metric becomes zero. \label{errors_wh_ham}}
\end{center}
\end{figure}
As in the previous subsection, the difference scheme may appear to be
convergent, but in fact it is not, see Figure \ref{errors_wh_ham}.  Frequency-dependent
exponential growth starting at very small values
 accumulates and causes the instability.  We have
exhaustively experimented with different values for the numerical
dissipation without being able to stabilize the code. Next
we show some plots to illustrate this. 
In figure \ref{dissipation1} we plot one of the Fourier components of
the numerical metric.  We increase the dissipation parameter $\tilde{\sigma}$,
starting from $\tilde{\sigma }= 0.02$,
and double it each time, while keeping the resolution and
Courant factor fixed. At the beginning the rate of the exponential
growth becomes smaller when one increases $\tilde {\sigma }$, though 
 it is never completely suppressed, which causes the
code to be non convergent. However,  when
 one reaches the value $\tilde{\sigma} = 0.32$ the instability is
even worse than adding less 
dissipation, and the same thing happens if one keeps on increasing
$\tilde{\sigma} $ beyond $0.32$. 
So we next  narrow the interval in which the dissipation is 
fine tuned, we start at $\tilde{\sigma} = 0.24$, and increase at intervals of
$0.01$. We find the same result, at 
$\tilde{\sigma} >0.25$ there is already too much dissipation and the situation
is worse. Fine tuning even more, we change $\tilde{\sigma} $ in intervals of 
$0.001$ around $0.250$,
but it is also found that for $\sigma >0.250$ the effect of more
dissipation is adverse, as also shown in figure
\ref{dissipation1}.

The fact that beyond $\tilde{\sigma} = 0.250$ the situation becomes
worse is in perfect  
agreement with the discrete analysis of \cite{convergence}. There we show that a necessary 
condition for the von Neumann condition to be satisfied is 
$\tilde{\sigma}\lambda \leq 1/8$. Here
the upper limit of $1/8$ corresponds to, precisely,
$\tilde{\sigma}=1/4$. Exceeding this value results in a violation of
the von Neumann condition; as explained in \cite{convergence}, when this
happens there is a numerical instability that grows explonentially
with the number of gridpoints 
(i.e. as in the CIP case), much
faster than when the von Neumann condition is satisfied (in which
case the growth goes as a power of the gridpoints). 

Finally, it is worthwhile to point out that we have also tried with 
smaller Courant factors, using, in particular,
values often used in numerical relativity, like $\lambda =0.20$ and 
$\lambda =0.25$, without ever being able to get a completely
convergent simulation.

\subsubsection{The CIP case}

\begin{figure}[ht] 
\begin{center} 
\includegraphics*[height=6cm]{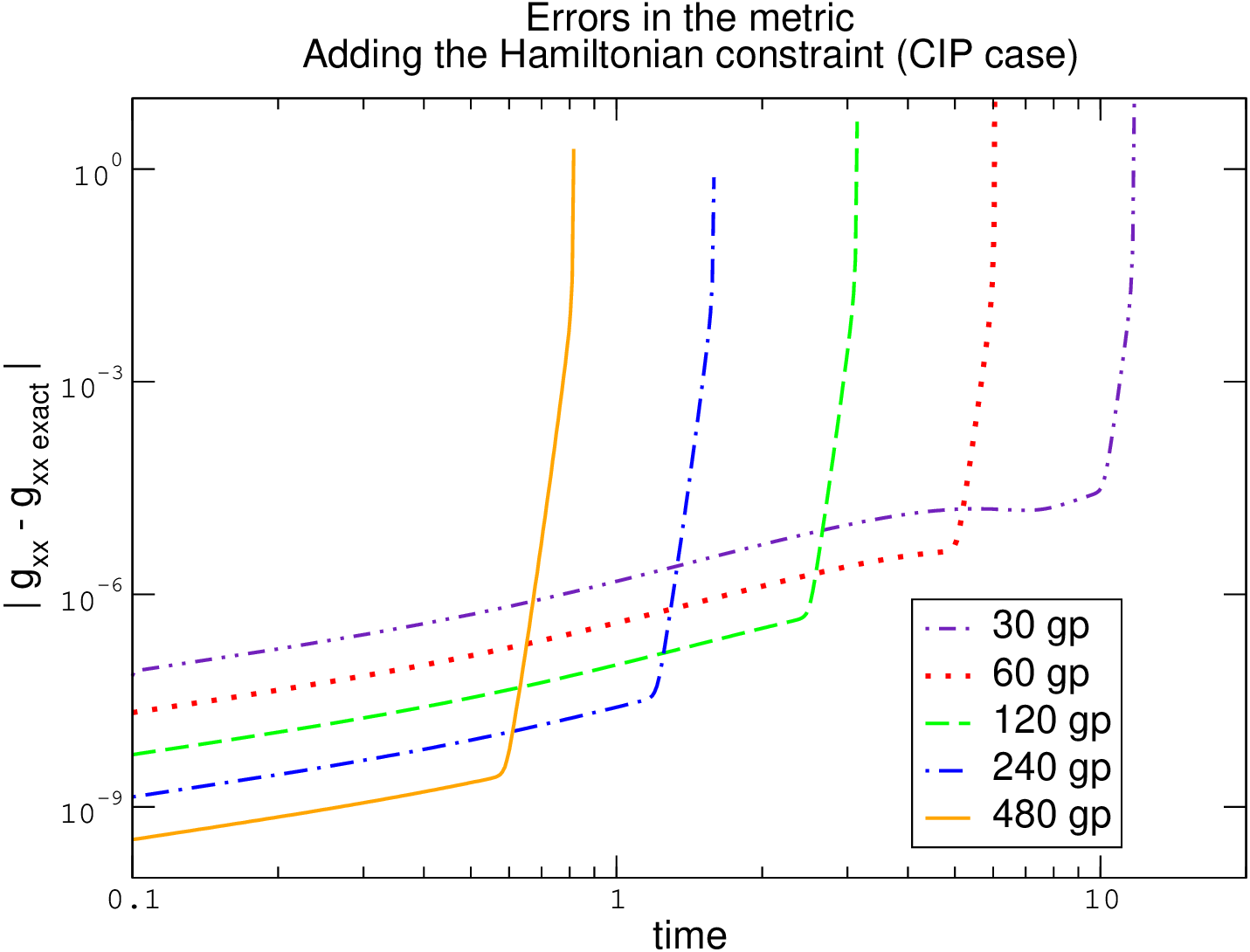} 
\caption{$L_2$ norm of the errors for the metric.  \label{errors_ip_constraint}}
\end{center}
\end{figure}

Finally here also use the parameters (\ref{par_sh}) with $\chi =1$, but now 
we take $\gamma = -79/42$, which implies $\lambda _2= -1$
and, thus, the system is CIP. The results are as expected. There is
exponential, frequency-dependent growth that makes the numerical
scheme unstable, see Figure \ref{errors_ip_constraint}.

\subsection{Other simulations}

\begin{figure}[ht] 
\begin{center} 
\includegraphics*[height=6cm]{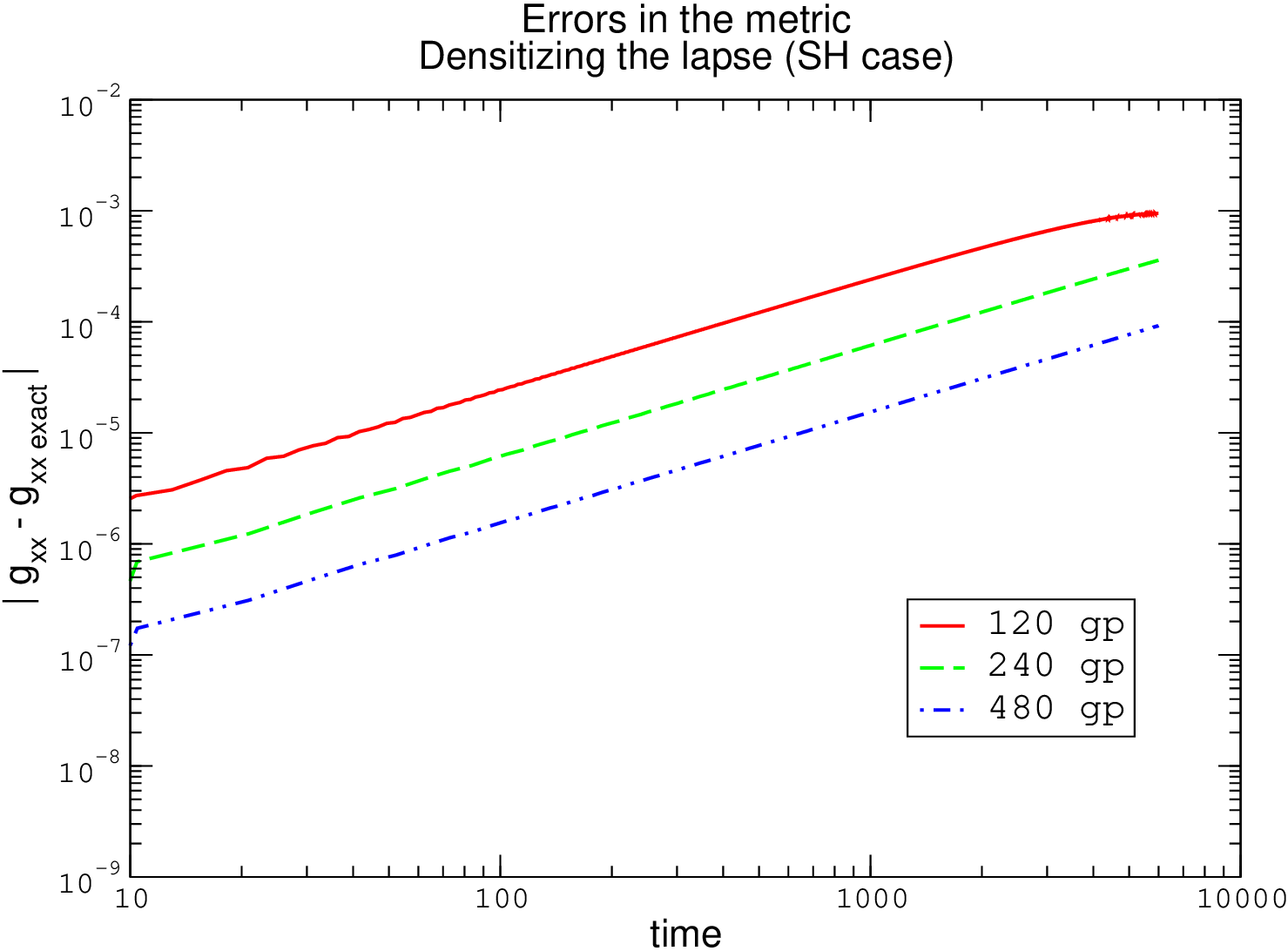} 
\includegraphics*[height=6cm]{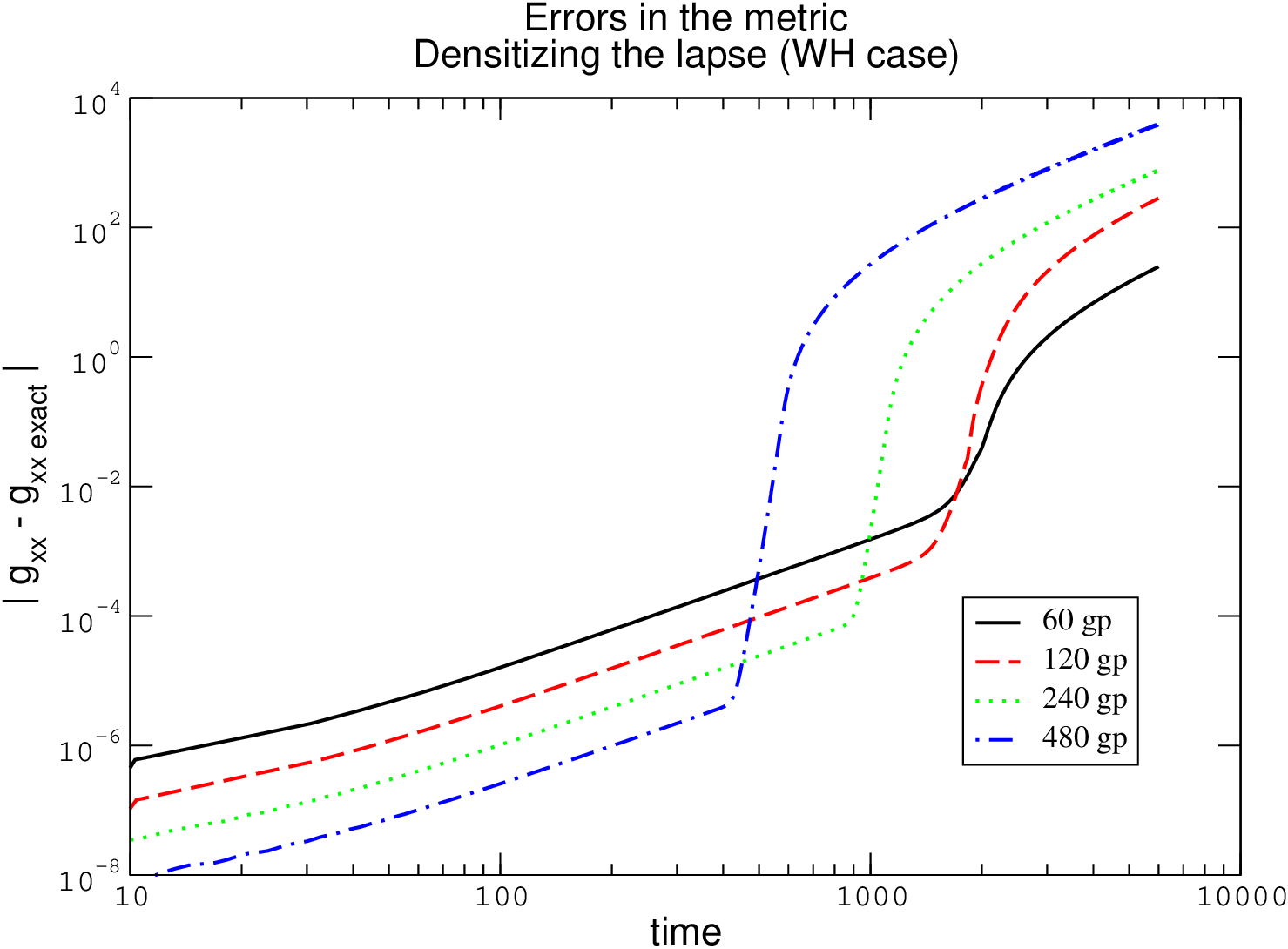} 
\caption{$L_2$ norm of the errors for the metric using ICN.  \label{errors_icn_dens}}
\end{center}
\end{figure}

Performing simulations with ICN instead of RK yields similar
results, as predicted in \cite{convergence}. Figure
\ref{errors_icn_dens} shows, 
for example, evolutions changing the
densitization of the lapse, as in the first subsection, but using ICN
with two iterations (counting this number as in \cite{teuk}). This is
the minimum number of iterations that yields a stable scheme for well
posed equations but, as shown in \cite{convergence}, it is unstable
for WH systems. The same values of Courant factor and dissipation as
above were used in these runs. 
We have also tried with other values of Courant factor and dissipation
parameter, 
finding similar results. We were able to confirm the lack of
convergence predicted in \cite{convergence}  in every WH or CIP formulation we
used, including the ADM
equations rewritten as first order equations in time
space. Lack of convergence with a 3D code, using the ADM equations
written as second order in space and ICN, for the same initial data
used here, has also been confirmed \cite{david}.

\section{Discussion}
\label{conclusions}

We have shown that a linearized analysis of the KST equations 
implies that flat space-time written in Cartesian coordinates
is a stable solution of the equations
if the parameters are chosen in such a way that the system is
strongly hyperbolic. No further restrictions on the parameters
are placed by this analysis. We have also integrated numerically
the KST equations and shown that the system cannot be made
stable (in the sense of Lax) and therefore convergent if the
parameters are not chosen in such a way that the system is 
strongly hyperbolic. No amount of artificial viscosity was
able to fix the problem.

The conclusions of Section \ref{kst} are, to some extent, similar to
those of Alcubierre {\it et al}., who find that, at the linearized level,
the advantages of the Baumgarte--Shapiro--Shibata--Nakamura \cite{bssn} with respect to 
the ADM formulation come from the fact that the
first one is `more' hyperbolic. However, since it is not `completely'
hyperbolic, an ill posed (zero speed, in the notation of \cite{alc1})
mode is still present in BSSN. The conformal traceless (CT) decomposition
is therefore introduced as a way of decoupling that mode. However, one
could, instead, get rid of that mode by using a different choice
of lapse. More explicitly, in the analysis of \cite{alc1} exact 
lapse is used, but it is not difficult to see that choosing, for
example, 
densitized lapse with, say, $\sigma =1/2$, gets rid of
the ill posed mode (indeed, certain version of BSSN with densitized
lapse has been recently used in a 3D evolution of a single black hole
\cite{pablo}). Moreover, as shown in \cite{our_bssn}, the result is
much stronger: even at the nonlinear level appropriately writing BSSN
with densitized lapse
results in a reduction (from first to second order in space) of
certain strongly or symmetric hyperbolic
formulations.

The lack of convergence in evolutions of WH or CIP systems with
schemes that are convergent for well posed equations seems to have
been overlooked in the past. If the von Neumann condition is satisfied, one
could be easily misled to think that the scheme is convergent,
especially if coarse resolutions are used, the evolution is short, or
few frequencies are present in the initial data. Here we have shown
that this lack of convergence does appear in concrete simulations,
even in very simple ones. These numerical experiments, together with
the theorem of reference \cite{convergence} should be enough evidence
to cast serious doubts on any simulation performed with evolution equations
that are weakly hyperbolic that are not accompanied by very detailed
convergence studies.

Summarizing, we do have analytical tools at our disposal to constrain
schemes and predict to a certain extent their behavior under numerical
evolution. In fact, the examples shown suggest that in the case of
linearized equations, the analytical tools are complete: one can a 
priori tell if a code will work or not. In the non-linear case further
tools will need to be developed to achieve the same status with 
respect to predicting the performance of a code before running it.

\section{Acknowledgments}

This work was supported in part by grants NSF-PHY-9800973, by funds from
the Horace Hearne Jr. Institute for Theoretical Physics, by the Swiss
National Science Foundation, and by Fundaci\'on Antorchas.  We thank many interesting
discussions with Larry Kidder, Saul Teukolsky, and Miguel 
Alcubierre. M.T. thanks Ed Seidel for hospitality
at AEI, where part of this work was done.




\begin{thebibliography}{10}

\bibitem{kst}
L.E. Kidder, M.A. Scheel, and S.A. Teukolsky,
Phys. Rev. D {\bf 64}, 064017 (2001).

\bibitem{convergence}
G. Calabrese, J. Pullin, O. Sarbach, and M. Tiglio, {\em Convergence
and stability in numerical relativity}, in preparation.

\bibitem{kreiss1}
H.O. Kreiss, J. Lorenz, {\em Initial-Boundary Value Problems and the Navier-Stokes Equations} (Academic Press, 1989)


\bibitem{richtmeyer}
R.D. Richtmyer and K.W. Morton, {\em Difference methods for initial-value problems}, 
(Krieger Publishing Company, Malabar, Florida, 1967)


\bibitem{fr}
S. Frittelli and O.A. Reula, Phys. Rev. Lett. {\bf
76}, 4667 (1996); S. D. Hern, Ph.D. thesis, University of Cambridge,
1999, gr-qc 0004036.

\bibitem{ec}
A. Anderson and J.W. York, Jr., Phys. Rev. Lett. {\bf 82}, 4384 (1999).

\bibitem{st} O. Sarbach and M. Tiglio, in preparation. 

\bibitem{thomas}
J.W. Thomas, {\em Numerical Partial Differential Equations, Vol I},
(Springer, New York, 1995),


\bibitem{teuk}
S.A. Teukolsky, Phys. Rev. D {\bf 61}, 087501 (2000).

\bibitem{david}
David Garrison, private communication.


\bibitem{bssn}
T.W. Baumgarte and S.L. Shapiro, Phys. Rev. D {\bf 59}, 024007
(1998); M. Shibata and T. Nakamura, Phys. Rev. D {\bf 52}, 5428
(1995).



\bibitem{alc1}
M. Alcubierre {\it et al}., Phys. Rev. D {\bf 62}, 124011 (2000).

\bibitem{pablo}
P. Laguna and D. Shoemaker, gr-qc/0202105. 

\bibitem{our_bssn} G. Calabrese, J. Pullin, O. Sarbach, and M. Tiglio,
gr-qc/0205064.




\end{thebibliography}
\end{document}